\documentclass[10pt,a4paper,twocolumn,french,american,english,aps,prb,amsmath]{revtex4}

\usepackage[T1]{fontenc}
\usepackage[latin9]{inputenc}
\usepackage{color}
\usepackage{babel}
\makeatletter
\addto\extrasfrench{%
   \providecommand{\fg}{\ifdim\lastskip>\z@\unskip\fi~\frqq}%
}

\makeatother
\usepackage{float}
\usepackage{bm}
\usepackage{amsmath}
\usepackage{amssymb}
\usepackage{graphicx}
\usepackage{esint}
\usepackage[unicode=true,pdfusetitle,
 bookmarks=true,bookmarksnumbered=false,bookmarksopen=false,
 breaklinks=false,pdfborder={0 0 1},backref=false,colorlinks=false]
 {hyperref}
\usepackage{breakurl}

\makeatletter

\providecommand{\tabularnewline}{\\}
\newcommand{\lyxdot}{.}

\@ifundefined{textcolor}{}
{%
 \definecolor{BLACK}{gray}{0}
 \definecolor{WHITE}{gray}{1}
 \definecolor{RED}{rgb}{1,0,0}
 \definecolor{GREEN}{rgb}{0,1,0}
 \definecolor{BLUE}{rgb}{0,0,1}
 \definecolor{CYAN}{cmyk}{1,0,0,0}
 \definecolor{MAGENTA}{cmyk}{0,1,0,0}
 \definecolor{YELLOW}{cmyk}{0,0,1,0}
}

\makeatother

\begin{document}

\title{Interplay between surface anisotropy and dipolar interactions in
an assembly of nanomagnets}

\author{Z. Sabsabi$^{1}$ }

\author{F. Vernay$^{1}$ }

\email{francois.vernay@univ-perp.fr}

\author{O. Iglesias$^{2}$ }

\email{oscariglesias@ub.edu}

\author{H. Kachkachi$^{1}$}

\affiliation{$^{1}$Laboratoire PROMES-CNRS UPR 8521 \& Université de Perpignan
Via Domitia, Rambla de la thermodynamique - Tecnosud, 66100 Perpignan,
France \\
$^{2}$Dept. de Física Fonamental and Institute of Nanoscience and
Nanotechnology (IN2UB), Facultat de Física, Universitat de Barcelona,
Av. Diagonal 647, 08028 Barcelona, Spain.}

\date{\today}
\begin{abstract}
We study the interplay between the effects of surface anisotropy and
dipolar interactions in monodisperse assemblies of nanomagnets with
oriented anisotropy. We derive asymptotic formulas for the assembly
magnetization taking account of temperature, applied field, core and
surface anisotropy, and dipolar inter-particle interactions. We find
that the interplay between surface anisotropy and dipolar interactions
is well described by the analytical expression of the assembly magnetization
derived here: the overall sign of the product of the two parameters
governing the surface and the dipolar contributions determines whether
intrinsic and collective terms compete or have synergistic effects
on the magnetization. This is illustrated by the magnetization curves
of $\gamma$-Fe$_{2}$O$_{3}$ nanoparticles assemblies in the low
concentration limit. 
\end{abstract}


\maketitle

\section{Introduction}

Assemblies of magnetic nanoparticles, deposited on a substrate or
embedded in a matrix, provide a very rich laboratory for various investigations,
experimental and theoretical, with stimulating challenges both for
fundamental research and practical applications.\cite{sharrock90ieee,johnson91jap,doretal97acp,batlab02jpd,tartajetal03jpd}
An issue of particular importance is that of the interplay between
intrinsic features of the nanomagnet pertaining to their finite-size
and boundary effects, and the collective effects induced by their
mutual interactions and their interactions with the hosting matrix
or substrate.\cite{cheetal95prb,respaudetal98prb,troetal00jmmm,kacgar05springer,kachkachi06neim} 

Experimental studies are numerous as they have concerned a variety
of parameters such as the production methods, particle sizes, shapes
and surface, matrices and substrates, degrees of concentration, organization
and aggregation.\cite{cheetal95prb,kodetal96prl,doretal97acp,gazetal97epl,kodamaetal97prl,mortro94prl,respaudetal98prb,petitetal98admat,shilovetal99prb,respaudetal99prb,sunmur99jap,troetal00jmmm,luisetal04jpcm,huangetal06epjb,leeetal11apl,russieretal12jap}
In particular, in some studies a subtle interplay was revealed between
the effects of the size distribution and the concentration of assemblies.
The results seem to hint to a kind of screening of the intrinsic effects
by the inter-particle interactions in dense assemblies. For example,
measurements of the magnetization at high fields performed on the
$\gamma$-Fe$_{2}$O$_{3}$ nanoparticles\cite{troetal00jmmm,ezzir98phd}
and on cobalt particles\cite{cheetal95prb} showed that the magnetization
is strongly influenced by surface effects. To be more explicit, Figs.
3 and 5 of Ref. \onlinecite{troetal00jmmm} together with Fig. 1 of
Ref. \onlinecite{kacetal00epjb} represent magnetization curves for
assemblies of maghemite nanoparticles with different concentrations.
One clearly sees that the $M(H)$ curves at different temperatures
present a rather different behavior as one compares dilute with concentrated
assemblies. Furthermore, measurements of the temperature $T_{\mathrm{max}}$
at the maximum of the zero-field-cooled (ZFC) magnetization as a function
of the applied magnetic field changes from a bell-like curve for dilute
assemblies into a monotonous decreasing function of the magnetic field
for concentrated systems.\cite{kacetal00jpcm} As of today the situation
has not changed very much regarding our understanding of the phenomena
observed in these experiments and many others. From a theoretical
point of view, the situation involving both acute surface effects
and long-ranged dipolar interactions has never been investigated in
detail so far mainly because of its tremendous complexities. In this
context, there are several contributions using either (semi)-analytical
approaches based on thermodynamic perturbation theory or mean-field
theory, or numerical studies such as the integration of the Fokker-Planck
equation or Monte Carlo simulations.\cite{kectro98prb,russieretal00ass,garciaoteroetal00prl,jongar01epl,bergor01jpcm,jongar01prb,kacaze05epjb,azekac07prb,kesserwanetal11prb,margarisetal12prb,ledueetal12jnn,russieretal12jap,landi13jap}

The dynamics of an assembly of magnetic nanoparticles, even weakly
interacting, is still another far more complex and challenging problem.
During the last two decades several models have been proposed to tackle
this issue with a special focus on the effects of DDI on the distribution
of the energy barriers and of the relaxation rates and the related
dynamical observables such as the ac susceptibility. One can find
in the literature many contradictory results as to whether DDI contribute
to enhance or to decrease the energy barrier of an interacting nanoparticle.\cite{doretal88jpc,mortro94prl,mamnak98jmmm,dormannetal99j3m,bergor01jpcm,jongar01prb,alliaetal01prb,mamiyaetal02prl,igllab04prb,chuchan05jap,masunagaetal09prb,ledueetal12jnn}
In Ref.  \onlinecite{jongar01epl} it was pointed out that these discrepancies
are mainly due to the fact that the discussions only focus on the
``static'' effect of DDI on the uniaxial energy potential and thus
overlook the fact that not only the energy landscape is important
but also how the magnetic moment evolves in it and how strongly its
motion is damped. Indeed, the presence of transverse components of
the local effective fields creates a saddle point in the uniaxial
potential barrier of the individual magnetic moments and this makes
the relaxation rate sensitive to damping. Unfortunately, dealing with
damping effects for studying the dynamics of an interacting assembly
of magnetic moments by \emph{e.g.} computing the relaxation rate(s)
of the assembly is a tremendous task which will still trigger many
investigations in the years to come. In this context, the relaxation
rate of the simplest nontrivial model of two magnetic moments coupled
by DDI has been computed in a study that has revealed several switching
mechanisms.\cite{francoetal11prb}

Real systems are assemblies of many-spin nanoparticles. On the other
hand, the investigation of surface effects requires an approach that
accounts for the internal structure of the nanoparticle regarded as
a nanocrystal of many atomic spins. Unfortunately, a system of interacting
many-spin particles is beyond the reach of any analytical calculation
and is of a horrendous difficulty for numerical simulations. A compromise
has been suggested in Ref.  \onlinecite{garkac03prl} where it is
shown that a many-spin particle can be modeled by a macroscopic magnetic
moment (the so called macrospin) evolving in an effective potential
that captures the main intrinsic features of the nanoparticle pertaining
to its size, shape, underlying lattice, and spin-spin interactions.
This potential comprises contributions that stem from the core and
surface anisotropies of the particle with coefficients whose sign
and magnitude depend on the intrinsic features.\cite{kacbon06prb,yanesetal07prb,kachkachi07j3m}
Therefore, this effective model then allows us to represent the magnetic
state of a nanoparticle by a single macroscopic magnetic moment while
taking account to some extent of the intrinsic properties. Consequently,
a dilute assembly of the so-represented nanoparticles provides a system
of weakly interacting magnetic moments each evolving in an effective
potential. Hence, one can deal with such a system using thermodynamic
perturbation theory with respect to the (weak) dipole-dipole interactions
(DDI) where the thermodynamic averages are computed with respect to
the Gibbs distribution for the effective potential energy.

Therefore, in the sequel we will consider and refer to the following
models: 
\begin{itemize}
\item One-spin problem (OSP), also known as the macrospin approximation:
each magnetic particle is modeled by a single magnetic moment, corresponding
to the net magnetic moment of the cluster, evolving in an energy potential
that comprises Zeeman and anisotropy contributions. Usually, the anisotropy
is uniaxial and incorporates the shape anisotropy as well. 
\item Effective one-spin problem (EOSP): in this model a many-spin magnetic
particle, inherently exhibiting surface anisotropy, can equivalently
be described by a single magnetic moment where surface effects are
taken into account by an effective potential. 
\end{itemize}
Then the task of the present work is to derive asymptotic analytical
expressions for the magnetization as a function of temperature and
applied field, which include contributions from the magnetic field,
core and surface anisotropies, and DDI. Upon varying the physical
parameters of the assembly such as its shape (oblate or prolate) and
concentration, it is possible to investigate the competition between
surface anisotropy and inter-particle interactions. In order to investigate
this competition without interference of other parameters such as
the volume and the easy axis distributions, we restrict our study
to mono-disperse assemblies with oriented anisotropy with all easy
axes pointing in the direction of the applied field. 

In Ref. \onlinecite{margarisetal12prb} the interplay between effects
of surface anisotropy and DDI was also studied using the same approach
with a computing method that consisted in a numerical calculation
of the integrals that yield the magnetization for individual particles.
A comparison with Monte Carlo calculations for arbitrary strength
of DDI was also presented. The present work is an important addition
in that it focuses on the analytical expressions for the magnetization.
More precisely, one of the major goals here is to provide practical
analytical expressions for the magnetization that take account of
the applied magnetic field, the (core) uniaxial anisotropy, the (surface)
cubic anisotropy, DDI, and temperature. This is indeed achieved in
limiting cases for the field, surface anisotropy and DDI. This analytical
tool allows us to discuss in more detail the competition between intrinsic
and collective effects, and is useful for a simple fitting procedure
of experimental data on dilute assemblies of rather small particles
with identified surface effects. The present results can also be helpful
in optimizing the physical parameters for new experimental studies
of nanoparticles assemblies in view of a better understanding of the
physical phenomena discussed above and eventually for the practical
applications in vogue nowadays, especially those which target a functionalization
of the particles surface.

After this Introduction, the paper is organized in 5 Sections: the
OSP and EOSP models and the corresponding general expression for the
magnetization are presented in Section \ref{sec:Presentation-of-the}.
Next, Section \ref{sec:Assembly-of-OSP} is devoted to the derivation
of low-field asymptotes, in the high- and low-anisotropy limits, for
the assembly magnetization. These analytical expressions provide a
link to previous studies and highlight the effect of the shape of
the assembly on the magnetization when solely the DDI are taken into
account. The competition between intrinsic and long-range terms in
the energy is then investigated in Section \ref{sec:EOSP-vs-DDI}
where the magnetization of the assembly within the EOSP approach is
derived. The analytical expressions computed in these last two Sections
are then discussed in detail in Section \ref{sec:Results-and-discussions}.
The paper ends with some concluding remarks and outlook.

\section{\label{sec:Presentation-of-the}Model and physical observables}

\subsection{\label{energy}Energy}

We consider an assembly of $\mathcal{N}$ ferromagnetic nanoparticles
each carrying a magnetic moment ${\bf m}_{i}=m_{i}{\bf s}_{i},\, i=1,\cdots,{\cal N}$
of magnitude $m_{i}$ and direction ${\bf s}_{i}$, with $\vert{\bf s}_{i}\vert=1$.
${\bf m}_{i}$ is then measured in terms of the Bohr magneton $\mu_{B}$,
\emph{i.e.} $m_{i}=n_{i}\mu_{B}$, and $n_{i}$ are either all equal
for monodisperse assemblies or chosen according to some distribution,
the so-called polydisperse assemblies. Each magnetic moment has a
uniaxial easy axis ${\bf e}_{i}$ and for an assembly these may be
either all directed along some reference axis leading to an oriented
assembly, or randomly distributed. The former situation is the one
that we consider here and we will refer to it as oriented anisotropy
(OA). Hence, the energy of a magnetic moment ${\bf m}_{i}$ interacting
with the whole assembly reads (after multiplying by $-\beta=-1/k_{B}T$)
\begin{equation}
\mathcal{E}_{i}={\cal E}_{i}^{(0)}+{\cal E}_{i}^{\mathrm{DDI}},\label{eq:DDIAssemblyEnergy}
\end{equation}
 where the first contribution 
\[
{\cal E}_{i}^{(0)}=x_{i}{\bf s}_{i}\cdot{\bf e}_{h}+\mathcal{A}\left({\bf s}_{i}\right)
\]
 is the energy of the free nanoparticle at site $i$ with the first
term being the Zeeman energy with the magnetic field having the verse
${\bf e}_{h}$. The second term is the anisotropy contribution where
$\mathcal{A}\left({\bf s}_{i}\right)$ is a function that depends
on the anisotropy model and is given by 
\begin{equation}
\mathcal{A}(\mathbf{s}_{i})=\left\{ \begin{array}{ll}
\sigma_{i}\left(\mathbf{s}_{i}\cdot\mathbf{e}_{i}\right)^{2}, & \quad\mathrm{OSP}\\
\\
\sigma_{i}\left[\left(\mathbf{s}_{i}\cdot\mathbf{e}_{i}\right)^{2}-\frac{\zeta}{2}\left(s_{i,x}^{4}+s_{i,y}^{4}+s_{i,z}^{4}\right)\right], & \quad\mathrm{EOPS}.
\end{array}\right.\label{eq:AnisotropyEnergy}
\end{equation}

As the main purpose of the present paper is to develop analytical
expressions, we restrict ourselves to textured samples where the uniaxial
anisotropies of Eq. (\ref{eq:AnisotropyEnergy}) remain along the
external magnetic field, i. e. ${\bf e}_{i}\parallel{\bf e}_{h}$.
Moreover, in the second line of Eq. (\ref{eq:AnisotropyEnergy}) we
assume the simplest situation where the three axes of the cubic anisotropy
coincide with the crystal axes and that one of them coincides with
the easy axis of the uniaxial anisotropy.

Within the EOSP model on-site surface anisotropy of the quadratic
kind, such as transverse or Néel's models, induces a quartic contribution
in the effective energy of the net magnetic moment, under the condition
that spin noncolinearities are not too strong.\cite{garkac03prl}
Later it was shown that\cite{yanesetal07prb,kachkachi07j3m} in fact
the effective energy of the many-spin particle is a polynomial in
the components of its net magnetic moment. This polynomial can be,
with a fairly good approximation, limited to a sum of a quadratic
and a quartic contribution with coefficients that change in sign and
magnitude with the intrinsic properties of the particle, namely its
size, shape, underlying lattice structure, and physical parameters
such as the exchange coupling, on-site core and surface anisotropy.
This has been checked numerically in Refs. \onlinecite{kacbon06prb,yanesetal07prb}. 

For convenience, in Eqs. (\ref{eq:DDIAssemblyEnergy}) and (\ref{eq:AnisotropyEnergy})
we have introduced the dimensionless parameters 
\begin{eqnarray}
x_{i} & \equiv & \frac{n_{i}\mu_{B}H}{k_{B}T}=n_{i}x_{0},\label{eq:dimless_params}\\
\sigma_{i} & \equiv & \frac{K_{2}V_{i}}{k_{B}T}=\frac{\left(\mu_{B}/M_{s}\right)K_{2}}{k_{B}T}n_{i}=\sigma_{0}n_{i},\nonumber \\
\zeta & \equiv & \frac{K_{4}}{K_{2}}.\nonumber 
\end{eqnarray}

${\bf H}=H{\bf e}_{h}$ is the external magnetic field, $V_{i}$ is
the volume of the nanoparticle, $K_{2}$ and $K_{4}$ are respectively
the quadratic and quartic anisotropy constants. Since we are considering
assemblies with moment instead of volume distribution, the volume
$V_{i}$ may be rewritten in terms of $n_{i}$ via the saturation
magnetization of the material per unit volume $M_{s}$, i.e. $V_{i}=m_{i}/M_{s}=(\mu_{B}/M_{s})n_{i}$.

In addition to the intrinsic contributions to the energy, the expression
in Eq. (\ref{eq:DDIAssemblyEnergy}) also includes a contribution
stemming from DDI which reads 
\begin{eqnarray}
{\cal E}_{i}^{\mathrm{DDI}} & = & \sum_{j<i}\xi_{ij}\frac{3({\bf s}_{i}\cdot{\bf e}_{ij})({\bf s}_{j}\cdot{\bf e}_{ij})-{\bf s}_{i}\cdot{\bf s}_{j}}{r_{ij}^{3}}\nonumber \\
 & = & \sum_{j<i}\xi_{ij}{\bf s}_{i}\cdot{\cal D}_{ij}\cdot{\bf s}_{j}\label{eq:osp_energy}
\end{eqnarray}
 with the corresponding dimensionless DDI coupling

\begin{equation}
\xi_{ij}=\left(\frac{\mu_{0}}{4\pi}\right)\left(\frac{\mu_{B}^{2}n_{i}n_{j}/a^{3}}{k_{B}T}\right).\label{eq:xi}
\end{equation}

${\cal D}_{ij}$ is the DDI tensor

\begin{equation}
{\cal D}_{ij}\equiv\frac{1}{r_{ij}^{3}}\left(3{\bf e}_{ij}{\bf e}_{ij}-1\right),\ {\rm with}\quad{\bf r}_{ij}={\bf r}_{i}-{\bf r}_{j},\ {\bf e}_{ij}=\frac{{\bf r}_{ij}}{r_{ij}}.\label{eq:DDITensor}
\end{equation}

For the sake of clarity of our discussions of the competition between
intrinsic and collective effects we will assume that the nanoparticles
are distributed on a stereotypical simple cubic super-lattice with
lattice parameter $a$. Therefore, the vector ${\bf r}_{ij}$ connects
the sites $i$ and $j$ and its magnitude is measured in units of
$a$.

It is useful to give some orders of magnitude for the DDI strength
in nanoparticle assemblies. For example, a cobalt atom carries a magnetic
moment of $n_{0}\simeq1.7$ Bohr magnetons. For two such atoms separated
by the atomic distance $a_{0}=0.3554\,\mathrm{nm}$, $\xi=\frac{1}{k_{B}T}\left(\frac{\mu_{0}}{4\pi}\right)\left(\mu_{B}^{2}n_{0}^{2}/a_{0}^{3}\right)\simeq0.004$
at 10K. A nanoparticle of diameter $D=3\,\mathrm{nm}$ contains circa
$n\simeq2172$ of such atoms. For two such particles separated by
a distance of $3D$ we have at the same temperature $\xi\simeq1.17$,
which is three orders of magnitude larger than for atoms.

Finally, let us relate the DDI parameter to the real assembly parameters,
especially the concentration of nanoparticles in a hosting matrix.
From the form of the DDI tensor $\mathcal{D}_{ij}$ in Eq. (\ref{eq:DDITensor})
it appears that DDI depend on two parameters: the inter-particle distance
and the relative orientation of the particles. The DDI strength can
equivalently be characterized by the parameter $\xi$ or the volume
concentration $C_{{\rm v}}$ which can be written as\cite{kectro98prb}

\selectlanguage{french}%
\begin{equation}
C_{{\rm v}}=p\left(\frac{k_{\mathrm{lat}}\pi}{6}\right)\left(\frac{D}{a}\right)^{3},\label{eq:Cv1}
\end{equation}
\foreignlanguage{english}{where $a$ is the lattice constant introduced
earlier, $D$ the particle diameter, and $k_{\mathrm{lat}}$ a constant
that depends on the lattice structure. For instance, for a simple
cubic lattice $k_{\mathrm{lat}}=1$. $p$ represents the occupancy
probability of a lattice site which for the present case is $p=1$.
Then, $C_{{\rm v}}$ and $\xi$ are related as follows
\[
\begin{array}{ccc}
\xi & = & \left(\frac{\mu_{0}}{4\pi}\right)\frac{\left(\mu_{B}n\right)^{2}}{k_{B}T}\frac{6}{\pi D^{3}}\times C_{{\rm v}}\\
\\
 & = & \frac{T_{0}}{T}\left(\frac{D}{a_{0}}\right)^{3}C_{{\rm v}}
\end{array}
\]
with $T_{0}$ and $a_{0}$ being constants corresponding respectively
to an arbitrary temperature and a length such that the ratio $T_{0}/a_{0}^{3}$
is given by 
\[
\frac{T_{0}}{a_{0}^{3}}=\frac{\mu_{0}\mu_{B}^{2}}{24k_{B}}\times\alpha^{2},
\]
where $\alpha$ represents the number of Bohr magnetons per unit volume.}

\selectlanguage{english}%
On the other hand, the presence of the unit vectors ${\bf e}_{ij}$
in Eq. (\ref{eq:osp_energy}) implies that the dipolar interaction
depends explicitly on the geometry of the sample. In order to illustrate
this effect we will consider two kinds of assemblies of $\mathcal{N}=2000$
particles: an oblate sample of dimensions ($20\times20\times5)$ and
a prolate sample ($10\times10\times20$).

\subsection{Magnetization}

The competition between intrinsic and collective effects in nanomagnet
assemblies affects most of the physical properties of the system inducing
a modification of various physical observables. In the present work,
we choose to focus on equilibrium properties and we therefore investigate
the behavior of the magnetization curves $M(H)$ taking account of
DDI. 

In Ref. \onlinecite{kacaze05epjb} it was shown that in a dilute assembly,
the magnetization of a nanocluster at site $i$ (weakly) interacting
with the other clusters of the assembly, is given (to first order
in $\xi$) by 
\begin{equation}
\left\langle s_{i}^{z}\right\rangle \simeq\langle s_{i}^{z}\rangle_{0}+\sum_{k=1}^{\mathcal{N}}\xi_{ik}\langle s_{k}^{z}\rangle_{0}A_{ki}\frac{\partial\langle s_{i}^{z}\rangle_{0}}{\partial x_{i}},\label{eq:DDIMag}
\end{equation}
where $A_{kl}={\bf e}_{h}\cdot{\cal D}_{kl}\cdot{\bf e}_{h}$. $\left\langle .\right\rangle $
is the statistical average of the projection on the field direction
of the particle's magnetic moment.

Eq. (\ref{eq:DDIMag}) was obtained for an external magnetic field
applied in the $z$ direction leading to $\langle s_{i}^{x,y}\rangle_{0}=0$.
It is only valid for a center-to-center inter-particle distance larger
than thrice the mean diameter of the nanoparticles.\cite{azekac07prb}
This implies that the magnetization of an interacting cluster is written
in terms of its ``free'' (with no DDI) magnetization and susceptibility,
with of course the contribution of the assembly ``lattice'' via
the lattice sum in Eq. (\ref{eq:DDIMag}). Therefore, one first has
to compute the magnetization of the free cluster. 

For an assembly of free particles it remains relatively straightforward
to derive an expression for the free magnetization, $m_{i}^{\left(0\right)}\equiv\langle s_{i}^{z}\rangle_{0}$,
for the specific case of uniaxial anisotropy corresponding to the
first line of Eq. (\ref{eq:AnisotropyEnergy}), \emph{i.e.} the OSP
model.\cite{garpal00acp,jongar01prb,kacaze05epjb} From this expression
of $m^{\left(0\right)}$, it is then easy to derive, using Eq. (\ref{eq:DDIMag}),
an analytical expression for the assembly magnetization that takes
account of the DDI in the dilute limit. Furthermore, as it will be
shown in Section \ref{sec:EOSP-vs-DDI}, an explicit expression for
$m^{\left(0\right)}$ allows one to derive a perturbative expression
for the magnetization for the EOSP by including the cubic anisotropy
term with the coefficient $\zeta$ as an expansion parameter.

As mentioned earlier, here we restrict ourselves to monodisperse assemblies
so as to investigate the interplay between intrinsic and collective
effects in pure form. Consequently, we set $x_{i}=x,\sigma_{i}=\sigma,\xi_{ij}=\xi$.
In this case, the magnetization of a (weakly) interacting particle
is given by Eq. (\ref{eq:DDIMag}) which now simplifies into the following
expression

\begin{equation}
\left\langle s^{z}\right\rangle \simeq m^{\left(0\right)}\left[1+\xi{\cal C}^{(0,0)}\frac{\partial m^{\left(0\right)}}{\partial x}\right].\label{eq:Mag_DDI}
\end{equation}
The lattice sum $\mathcal{C^{\mathrm{(0,0)}}}$ is in fact the first
of a hierarchy of lattice sums.\cite{kacaze05epjb} For large samples,
$\mathcal{C}^{\left(0,0\right)}$may be rewritten in term of the demagnetizing
factor $D_{z}$ along $z$\cite{jongar01prb}
\[
\mathcal{C}^{\left(0,0\right)}=-4\pi\left(D_{z}-\frac{1}{3}\right).
\]

In the continuum limit and for box-shaped samples of semi-axes $a,$$b,$
$c$, $D_{z}$ can itself be expressed as follows 
\[
D_{z}=\frac{abc}{2}{\displaystyle \int_{0}^{\infty}\frac{ds}{\left(c^{2}+s\right)\sqrt{\left(a^{2}+s\right)\left(b^{2}+s\right)\left(c^{2}+s\right)}}}.
\]
This form of $\mathcal{C}^{\left(0,0\right)}$ allows one to have
a direct evaluation for large samples of the demagnetizing factor
$D_{z}$ known for different sample aspect ratios. \cite{AharoniJAP1998}

It then turns out that the relevant DDI parameter, to this order of
approximation, is in fact 
\[
\tilde{\xi}\equiv\xi\mathcal{C}^{\left(0,0\right)}=\frac{\xi}{\mathcal{N}}\sum_{i,j=1,i\neq j}^{{\cal N}}A_{ij}.
\]

Next, the longitudinal susceptibility $\chi_{\parallel}^{\left(0\right)}=\partial m^{\left(0\right)}/\partial x$
is given by {[}see \emph{e.g.} Ref. \onlinecite{garpal00acp} for
a review{]}

\[
\frac{\partial m^{\left(0\right)}}{\partial x}=\frac{1+2S_{2}}{3}-\left(m^{\left(0\right)}\right)^{2}=a_{0}^{\left(1\right)}-\left(m^{\left(0\right)}\right)^{2}
\]
 and thereby we obtain the approximate expression for the magnetization
of a particle taking account of DDI with the other particles in the
assembly 
\begin{equation}
\left\langle s^{z}\right\rangle \simeq m^{\left(0\right)}\left[1+\tilde{\xi}\left(a_{0}^{\left(1\right)}-\left(m^{\left(0\right)}\right)^{2}\right)\right].\label{eq:Mag_DDI_explicit}
\end{equation}

The free particle magnetization $m^{\left(0\right)}$ can be computed
in various ways, either numerically or analytically. Our major concern
in this work is to provide ready-to-use analytical expressions for
the magnetization of an assembly, including DDI and surface effects.
For this purpose, we adopt an analytical approach with the understanding
that this can only be performed in some limiting cases of the applied
field and quartic contribution to the particle's anisotropy, \emph{i.e.}
low and high fields and/or small $\zeta$.

Obviously, in the absence of any interaction and anisotropy, or at
high temperature, the magnetization is described by the Langevin function

\begin{equation}
\langle s^{z}\rangle_{0}\left(\sigma=0,\xi=0\right)=\mathcal{L}\left(\frac{\mu_{0}HM_{S}}{k_{B}T}\right),\label{eq:Langevin}
\end{equation}
where $\mu_{0}$ is the vacuum permeability introduced so that $\mu_{0}H$
can be measured in Tesla.

\section{\label{sec:Assembly-of-OSP}Assembly of OSP nanoclusters : oriented
uniaxial anisotropy vs DDI}

In the case of a nanocluster with effective uniaxial anisotropy in
a longitudinal magnetic field, \emph{i.e.} $\mathbf{e}_{i}\parallel\mathbf{e}_{h}\parallel\mathbf{e}_{z}$,
the energy reads (dropping the particle's index $i$) 
\[
{\cal E}^{(0)}=\sigma s_{z}^{2}+xs_{z}.
\]
Then, we introduce the ``free\textquotedbl{} probability distribution
\begin{equation}
\mathcal{P}_{0}\left(z\right)=\dfrac{1}{Z_{\parallel}^{\left(0\right)}}e^{{\cal E}^{(0)}},\ Z_{\parallel}^{\left(0\right)}\left(\sigma,x\right)=\int_{-1}^{1}ds_{z}\, e^{{\cal E}^{(0)}}\equiv\int d\omega^{\left(0\right)}.\label{eq:FreePF}
\end{equation}
 The free partition function $Z_{\parallel}^{^{(0)}}$ may then be
rewritten in terms of the Dawson integral $D\left(x\right)=e^{-x^{2}}\int_{0}^{x}dt\, e^{t^{2}}$
as\cite{garpal00acp} 
\[
Z_{\parallel}^{\left(0\right)}\left(\sigma,x\right)=\dfrac{e^{\sigma}}{\sqrt{\sigma}}\left[e^{x}D(\sqrt{\sigma_{+}})+e^{-x}D(\sqrt{\sigma_{-}})\right]
\]
where the reduced field $h=x/2\sigma$ and energy barriers $\sigma_{\pm}\equiv\sigma\left(1\pm h\right)^{2}$
have been introduced. 

The magnetization in the presence of anisotropy and longitudinal field
is then given by
\begin{equation}
\langle s^{z}\rangle_{0}\left(\sigma\neq0,\xi=0\right)=\dfrac{e^{\sigma}}{\sigma Z_{\parallel}^{\left(0\right)}}\sinh x-h=\mathcal{C}_{1}.\label{eq:MzLongField}
\end{equation}

$\mathcal{C}_{1}$ is defined in the Appendix.

Now, before using this expression in Eq. (\ref{eq:Mag_DDI_explicit})
to derive the corresponding expression for the assembly magnetization,
we recall some asymptotes for the magnetization of the free cluster
in various regimes of the anisotropy and applied field. The details
of the calculations and more asymptotes can be found in Ref. \onlinecite{garpal00acp}. 

First, from Eq. (\ref{eq:MzLongField}) one derives low- and high-anisotropy
asymptotes

\[
m^{\left(0\right)}\simeq\left\{ \begin{array}{lll}
\mathcal{L}\left(x\right)+\frac{2}{x}\left[\mathcal{L}^{2}\left(x\right)-\left(1-\frac{3}{x}\mathcal{L}\left(x\right)\right)\right]\sigma, &  & \sigma\ll1,\\
\\
\tanh x\left[1-\frac{1}{2\sigma}\left(1+\frac{2x}{\sinh\left(2x\right)}\right)\right], &  & \sigma\gg1.
\end{array}\right.
\]
and similarly for the longitudinal susceptibility we get

\begin{widetext} 
\[
\frac{\partial m^{\left(0\right)}}{\partial x}\simeq\left\{ \begin{array}{lll}
\frac{1}{3}-\mathcal{L}^{2}\left(x\right)+4\left\{ \frac{1}{45}-\frac{\mathcal{L}\left(x\right)}{x}\left[\mathcal{L}^{2}\left(x\right)-\left(1-\frac{3}{x}\mathcal{L}\left(x\right)\right)\right]\right\} \sigma, &  & \sigma\ll1,\\
\\
\left(1-\tanh^{2}x\right)-\left[1-\tanh^{2}x\left(1+\frac{2x}{\sinh\left(2x\right)}\right)\right]\frac{1}{\sigma}, &  & \sigma\gg1.
\end{array}\right.
\]
 \end{widetext}

Since the present investigation focuses on the collective equilibrium
behavior in the low-temperature limit, we note in passing that typical
physical parameters of (metallic or oxide) nanoparticles are such
that $\sigma\gg1$. 

Next, for the assembly we use Eq. (\ref{eq:Mag_DDI_explicit}) to
derive the low- and high-field asymptotes in both anisotropy regimes.
For low fields, we use the low-field expansion of the Langevin function
($\mathcal{L}\left(x\right)\simeq\frac{x}{3}-\frac{x^{3}}{45}$) to
get the low-field asymptotes for the assembly magnetization 
\[
\left\langle s^{z}\right\rangle _{\mathrm{LF}}\simeq\left\{ \begin{array}{lll}
m^{\left(0\right)}\left[1+\tilde{\xi}\left(\frac{1}{3}+\frac{4}{45}\sigma\right)\right], &  & \sigma\ll1,\\
\\
m^{\left(0\right)}\left[1+\tilde{\xi}\left(1-\frac{1}{\sigma}\right)\right], &  & \sigma\gg1.
\end{array}\right.
\]

For high fields we have $\mathcal{L}\left(x\right)\simeq1-\frac{1}{x}$
which leads to the following asymptotes

\[
\left\langle s^{z}\right\rangle _{\mathrm{HF}}\simeq\left\{ \begin{array}{lll}
m^{\left(0\right)}\left[1+\tilde{\xi}\left\{ \left(-\frac{2}{3}+\frac{4}{45}\sigma\right)+\frac{2}{x}\right\} \right], &  & \sigma\ll1,\\
\\
m^{\left(0\right)}\left[1+\tilde{\xi}\ \mathcal{O}(\frac{1}{\sigma^{2}})\right], &  & \sigma\gg1.
\end{array}\right.
\]

The magnetization and thermal averages of higher-order moments are
given in the Appendix. 

\begin{figure}[H]
\begin{centering}
\begin{center}\includegraphics[width=1\columnwidth]{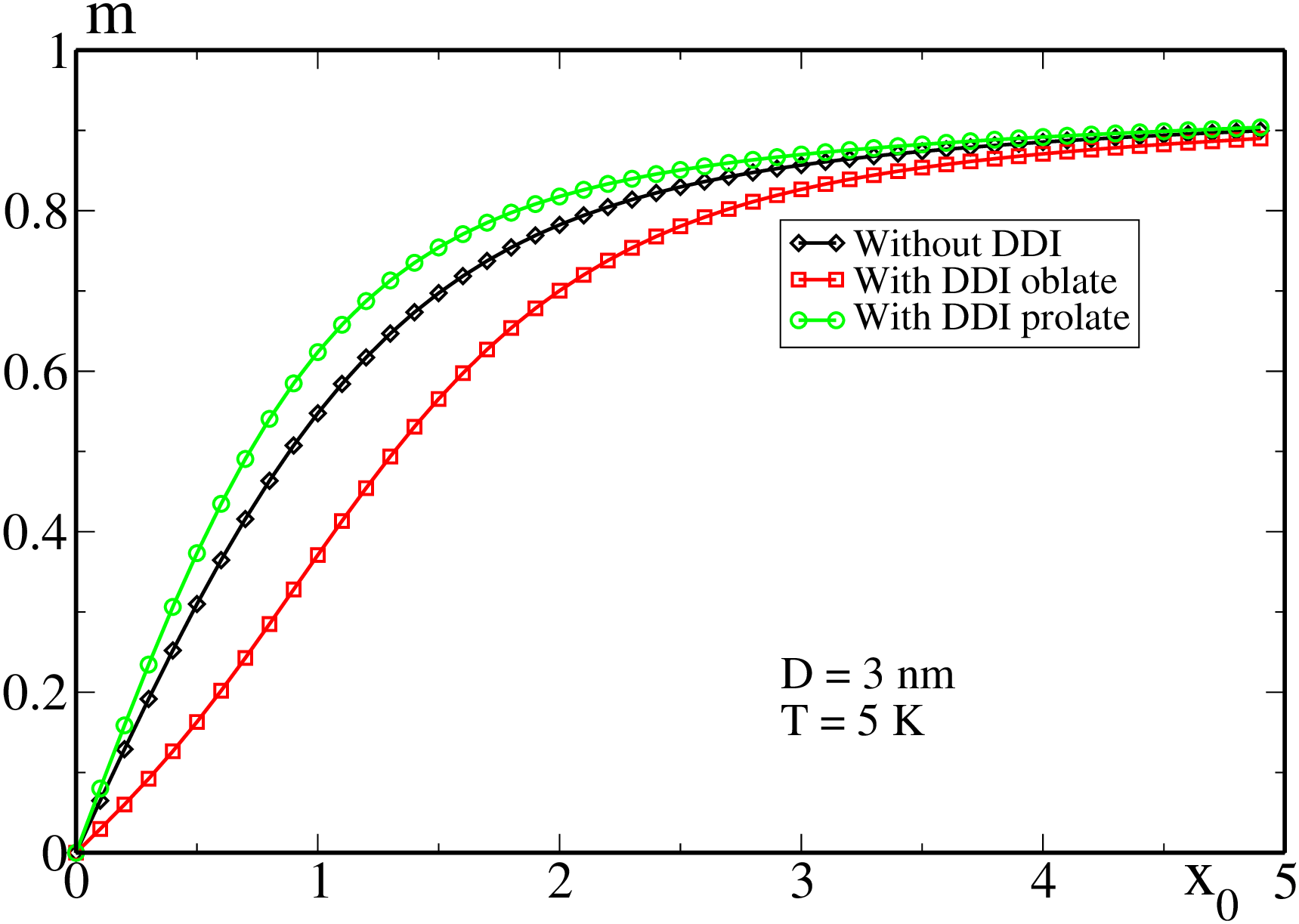}\end{center}
\par\end{centering}

\caption{\label{fig:MagExact-vs-DDICorrected}(color online) Reduced magnetization
of two assemblies of equivalent sizes, of prolate and oblate shape,
with ($\xi=0.18$) or without ($\xi=0$) DDI.}
\end{figure}

While the previous low-field and high-field asymptotes for the magnetization
are useful for various estimations, one can of course use the exact
semi-analytical expression of Eq. (\ref{eq:MzLongField}), insert
it in Eq. (\ref{eq:Mag_DDI_explicit}) and obtain the magnetization
for the whole range of the applied field. The result of this procedure
is shown in Fig.\ref{fig:MagExact-vs-DDICorrected}.

Here the effect of DDI and of the sample's shape is obvious. As is
well known, see \emph{e.g. }Ref. \onlinecite{kacaze05epjb} and references
therein, DDI are anisotropic interactions and thus contribute to the
effective anisotropy. Since the anisotropy is uniaxial and oriented,
\emph{i.e.} with a common easy axis, its effect leads to a magnetization
enhancement. On the contrary, the DDI effect depends on the sign of
$\tilde{\xi}$ (or more precisely that of $\mathcal{C}^{\left(0,0\right)}$),
which is directly related to the sample's shape. For instance, in
the case of oblate samples $\mathcal{C}^{\left(0,0\right)}<0$ leading
to a reduction of the magnetization, while for prolate samples $\mathcal{C}^{\left(0,0\right)}>0$
and thereby DDI contribute to the enhancement of the assembly's magnetization.
Consequently, for oblate samples the (oriented) uniaxial anisotropy
and DDI have opposite effects while for prolate samples they play
concomitant roles.

The integral of $Z_{\parallel}^{\left(0\right)}$ in Eq. (\ref{eq:MzLongField})
may also be extended to include the quartic contribution to the energy
potential and thereby deal with surface anisotropy and its competition
with DDI. This will be shown in the next section. The numerical calculations
of the integral of $Z_{\parallel}^{\left(0\right)}$ were done in
Ref. \onlinecite{margarisetal12prb}.

\section{\label{sec:EOSP-vs-DDI}Assembly of EOSP clusters: surface anisotropy
vs DDI}

The aim of this section is to deal with (weakly) interacting assemblies
of EOSP particles and investigate the interplay between surface anisotropy
and DDI. This means that in addition to the uniaxial anisotropy and
Zeeman contributions, the free-particle energy also includes a quartic
contribution in the components of its net magnetic moment according
to the second line of Eq. (\ref{eq:AnisotropyEnergy}). This effective
potential comprises a cubic anisotropy contribution, with the coefficient
$\zeta$, \citep{garkac03prl,kacbon06prb,kachkachi07j3m,yanesetal07prb}
that accounts for the surface anisotropy. The latter is initially
modeled in the many-spin description of a nanocluster with the help
of Néel's model for atomic spins.

Making use of the condition $\left\Vert \bm{s}\right\Vert =1$ the
cubic-anisotropy energy that is usually written as 

\selectlanguage{french}%
\begin{equation}
E_{CA}=-K_{4}V(s_{x}^{2}s_{y}^{2}+s_{z}^{2}s_{y}^{2}+s_{x}^{2}s_{z}^{2})\label{eq:energie-aniscub-0}
\end{equation}
\foreignlanguage{english}{can be recast in the more compact expression
(upon dropping an irrelevant constant) $E_{CA}=\frac{K_{4}V}{2}{\displaystyle \sum_{\alpha=x,y,z}}s_{\alpha}^{4}$.
Then, using the notation in Eq. (\ref{eq:AnisotropyEnergy}) we define
the (dimensionless) energy $\mathcal{E}_{CA}=\frac{\zeta}{2}\sum_{\alpha=x,y,z}s_{\alpha}^{4}$.
For later discussion we recall that depending on the sign of $\zeta$
there are different easy axes:}
\selectlanguage{english}%
\begin{itemize}
\item If $\zeta>0$, there are 8 minima, 12 saddle points, and 6 maxima.
The easy axes lie along the main diagonal of the cube,
\item If $\zeta<0$, there are 6 minima, 12 saddle points, and 8 maxima.
The easy axes are along $\left(Ox\right)$, $\left(Oy\right)$ and
$\left(Oz\right)$.
\end{itemize}

\subsection{Surface effects in the absence of DDI}

First we would like to highlight the effect of surface anisotropy
on the magnetization curves. So here we compute the magnetization
without the DDI contribution (\emph{i.e.} $\xi=0$). Hence, the partition
function is given by 
\[
Z=\int d\varphi d\omega^{\left(0\right)}e^{-\frac{\sigma\zeta}{2}\sum_{\alpha=x,y,z}s_{\text{\ensuremath{\alpha}}}^{4}}.
\]
Then, we assume that the cubic anisotropy is small. Indeed, the condition
of validity for the EOSP model \citep{garkac03prl,kacbon06prb,yanesetal07prb,kachkachi07j3m}
obtained for a nanocluster with an SC crystal lattice is $\zeta=K_{4}/K_{2}\lesssim1/4$
($\zeta\lesssim0.35$ for an FCC lattice). In this case, the spin
noncolinearities induced by surface anisotropy are supposed not to
be too strong and thereby the anisotropy energy minima are mainly
determined by the uniaxial contribution, whereas the cubic contribution
only introduces saddle points. This leads to larger relaxation rates\citep{dejardinetal08jpd}
but does not affect the physical properties at equilibrium. 

Consequently, it is quite legitimate to expand the partition function
$Z$ in terms of $\zeta$ leading to

\begin{equation}
Z\simeq Z_{\parallel}^{\left(0\right)}-\frac{\sigma\zeta}{2}\left(Z_{\parallel}^{\left(2\right)}+Z_{\perp}^{\left(2\right)}\right)\label{eq:PFExpanded}
\end{equation}
where the ``longitudinal'' partition functions $Z_{\parallel}^{\left(0\right)}$
and $Z_{\parallel}^{\left(2\right)}$ are defined in Eqs. (\ref{eq:FreePF},
\ref{eq:FreePFn}), while the transverse component $Z_{\perp}^{\left(2\right)}$
is given by

\[
Z_{\perp}^{\left(2\right)}=\int d\varphi\left(s_{x}^{4}+s_{y}^{4}\right)d\omega^{\left(0\right)}.
\]

Next, by symmetry we have $\left\langle s_{x}^{4}\right\rangle =\left\langle s_{y}^{4}\right\rangle =\frac{1}{4}\left\langle \left(1-2s_{z}^{2}+s_{z}^{4}\right)\right\rangle =\frac{1}{4}\left\langle \left(1-s_{z}^{2}\right)^{2}\right\rangle $
and consequently $Z_{\perp}^{\left(2\right)}$ becomes
\[
Z_{\perp}^{\left(2\right)}=\frac{1}{2}Z_{0}^{\left(0\right)}-Z_{\parallel}^{\left(1\right)}+\frac{1}{2}Z_{\parallel}^{\left(2\right)}.
\]

Inserting this result back into Eq. (\ref{eq:PFExpanded}) and using
Eq. (\ref{eq:FreePFn}), we rewrite the partition function $Z$ in
terms of $Z_{\parallel}^{\left(0\right)}$ and its derivatives with
respect to $\sigma$ ($\partial_{\sigma}^{n}Z_{\parallel}^{\left(0\right)}$)

\begin{equation}
Z\simeq Z_{\parallel}^{\left(0\right)}+\frac{\sigma\zeta}{2}\left\{ \left(\partial_{\sigma}Z_{\parallel}^{\left(0\right)}\right)-\frac{1}{2}\left[Z_{\parallel}^{\left(0\right)}+3\left(\partial_{\sigma}^{2}Z_{\parallel}^{\left(0\right)}\right)\right]\right\} .\label{eq:partition_Z_Z0}
\end{equation}

This can further be rewritten in terms of Legendre polynomials as
follows 
\begin{equation}
Z\simeq Z_{\parallel}^{\left(0\right)}\left\{ 1-\sigma\zeta\left[\frac{7}{30}+\frac{2}{21}\mathcal{C}_{2}+\frac{6}{35}\mathcal{C}_{4}\right]\right\} \label{eq:Partition_Z_Effect-surfe}
\end{equation}
 with $\mathcal{C}_{2}$ and $\mathcal{C}_{4}$ being given in Eq.
 (\ref{eq:12Moments}).

\begin{figure}[H]
\centering{} \includegraphics[width=1\columnwidth]{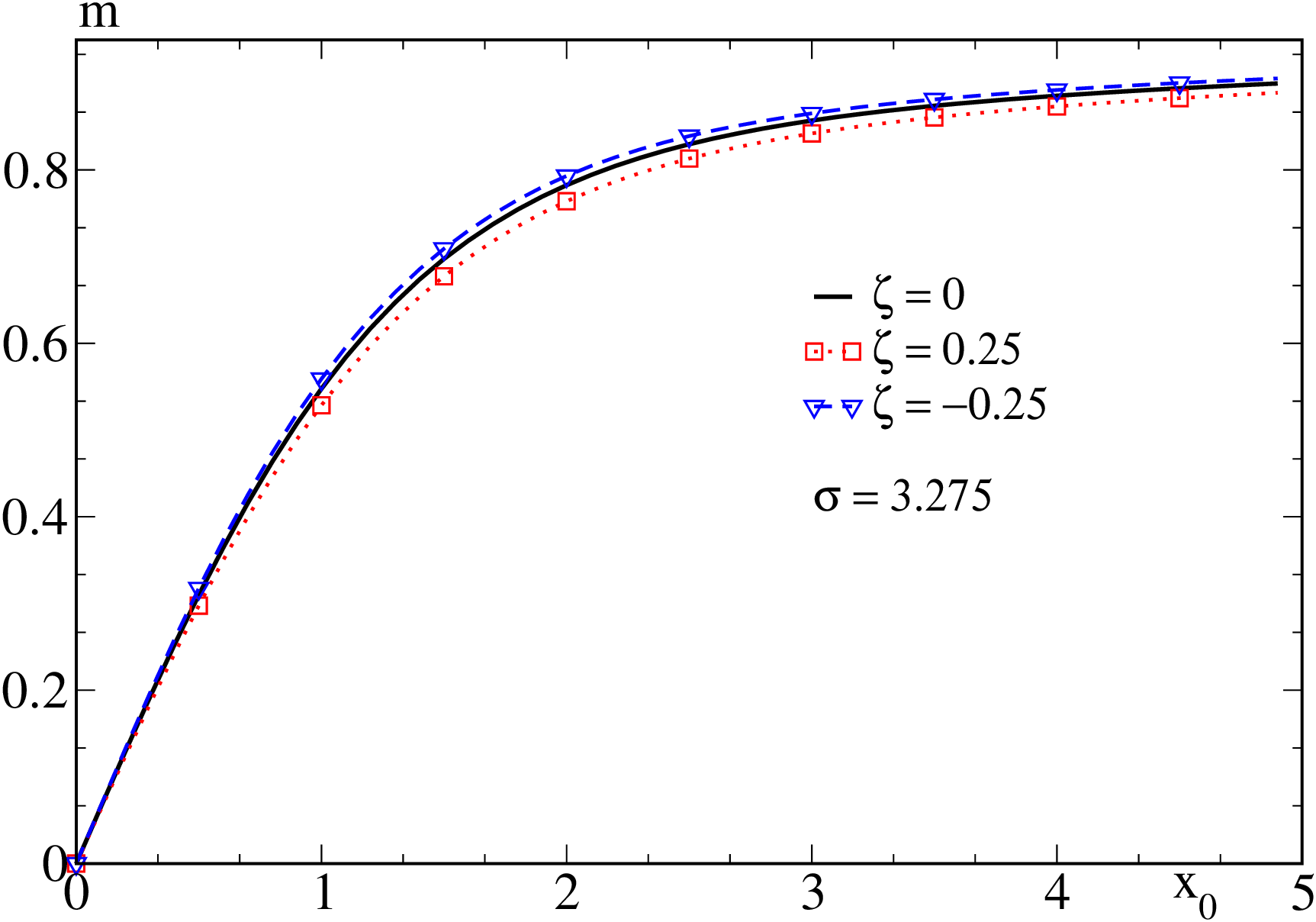}
\caption{(color online) Magnetization as a function of the (dimensionless)
field $x_{0}=\frac{\mu_{B}H}{k_{B}T}$ at temperature $T=5K$ and
various cubic anisotropy values $\zeta=K_{4}/K_{2}$\foreignlanguage{american}{\label{fig:Magnetization-axial+cubic}.}}
\end{figure}

Now the magnetization can be computed from the partition function
using $m^{\left(0\right)}=\frac{1}{Z}\partial Z/\partial x$. Unfortunately,
this leads to a cumbersome expression that we omit here. Representative
magnetization curves for two signs of the cubic anisotropy are shown
in Fig. \ref{fig:Magnetization-axial+cubic}. We see that for negative
(positive) $\zeta$ the nanoparticle assembly is respectively easier
(harder) to magnetize than in the case with only uniaxial anisotropy.

Similarly to what was done in Section \ref{sec:Assembly-of-OSP} for
the OSP model, one can establish for the EOSP model analytical asymptotes
in various field and anisotropy regimes. In the limit $\sigma\gg1$
the partition function $Z_{\parallel}^{\left(0\right)}$ reads {[}see
Eq. (2.74) of Ref. \onlinecite{garpal00acp} for the case $\zeta=0$
and arbitrary field{]}
\[
Z_{\parallel}^{\left(0\right)}=\frac{e^{\sigma}}{\sigma}\cosh x\left\{ \begin{array}{l}
1+\frac{1}{2\sigma}\left(1-x\tanh x\right)+\\
\\
\frac{1}{4\sigma^{2}}\left[\left(3+x^{2}\right)-3x\tanh x\right]
\end{array}\right\} .
\]

Then, substituting the latter in Eq. (\ref{eq:partition_Z_Z0}) and
performing a double expansion, with respect to $x$ for low fields
and $1/\sigma$ for high anisotropy, leads to the following expression
for the magnetization within the EOSP approach
\begin{align}
m^{\left(0\right)}\left(x,\sigma,\zeta\right) & \simeq\left(1-\frac{1}{\sigma}\right)x-\left(1-\frac{2}{\sigma}\right)\frac{x^{3}}{3}\nonumber \\
 & +\frac{\zeta}{\sigma}\left[-\left(1-\frac{2}{\sigma}\right)x+\left(2-\frac{5}{\sigma}\right)\frac{x^{3}}{3}\right].\label{eq:MagnetizationEOSP}
\end{align}

This can also be rewritten in the form $m^{\left(0\right)}\simeq\chi^{\left(1\right)}x+\chi^{\left(3\right)}x^{3}$
where one can easily identify the EOSP corrections to the linear and
cubic susceptibilities (in the limit of high-anisotropy barrier) corrected
by surface anisotropy 
\begin{eqnarray}
\chi^{\left(1\right)} & \simeq & \left(1-\frac{1}{\sigma}\right)+\frac{\zeta}{\sigma}\left(-1+\frac{2}{\sigma}\right),\nonumber \\
\chi^{\left(3\right)} & \simeq & \frac{1}{3}\left[\left(-1+\frac{2}{\sigma}\right)+\frac{\zeta}{\sigma}\left(2-\frac{5}{\sigma}\right)\right].\label{eq:MagnetizationEOSP2}
\end{eqnarray}

The competition between the uniaxial and cubic anisotropy contributions
can be understood as follows. As discussed earlier {[}see also Ref.
\onlinecite{margarisetal12prb} {]}, for $\zeta>0$ the energy minima
of the cubic contribution are along the cube diagonals $\left[\pm1,\pm1,\pm1\right]$
while for $\zeta<0$ they are along the cube edges $\left[1,0,0\right],\left[0,1,0\right],\left[0,0,1\right]$.
Hence, the uniaxial anisotropy with an easy axis along the $z$ direction,
\emph{i.e.} $\left[0,0,1\right]$, competes with the cubic anisotropy
when $\zeta>0$ whereas the two anisotropies have a concomitant effect
when $\zeta<0$. In the former case, the particle's magnetic moment
at equilibrium takes an intermediate direction between the $z$ axis
and the cube diagonal. Hence, as $\zeta$ increases the magnetic moment
gradually rotates away from the $z$ axis and thereby its statistical
average, or the magnetization, decreases. In the case of negative
$\zeta$ the two anisotropies cooperate to quickly drive the magnetization
towards saturation.

\subsection{Surface effects in the presence of DDI}

Finally, we derive the asymptotic expressions for the magnetization
taking account of both surface anisotropy and DDI, in addition of
course to the contributions from the uniaxial anisotropy and magnetic
field. Accordingly, using the asymptotic expression (\ref{eq:MagnetizationEOSP2})
in Eq. (\ref{eq:Mag_DDI_explicit}) leads to

\begin{align}
m\left(x,\sigma,\zeta,\tilde{\xi}\right) & \simeq\tilde{\chi}^{\left(1\right)}x+\tilde{\chi}^{\left(3\right)}x^{3}\label{eq:MagEOSPvsDDI1}
\end{align}
where

\begin{align}
\tilde{\chi}^{\left(1\right)} & \simeq\chi^{\left(1\right)}+\tilde{\xi}\left[1-\frac{2}{\sigma}-2\left(1-\frac{3}{\sigma}\right)\frac{\zeta}{\sigma}\right],\label{eq:MagEOSPvsDDI2}\\
\nonumber \\
\tilde{\chi}^{\left(3\right)} & \simeq\chi^{\left(3\right)}-\frac{4}{3}\tilde{\xi}\left[\left(1-\frac{3}{\sigma}\right)-\frac{3\zeta}{\sigma}\right].\nonumber 
\end{align}
are the linear and cubic susceptibilities of Eq. (\ref{eq:MagnetizationEOSP2})
corrected by DDI.

This is an important result of this work that is directly related
to its title. Indeed, this asymptotic expression allows us to figure
out how surface anisotropy competes with DDI. More precisely, the
sign of surface contribution with intensity $\zeta$ plays an important
role in the magnetization curve. Yet, as it couples to the DDI $\tilde{\xi}$
parameter, which contains information on the sample's shape, it is
the overall sign of $\tilde{\xi}\zeta$ that determines whether there
is a competition between surface and DDI effects or if the changes
in magnetization induced by the intrinsic and collective contributions
have the same tendency. The answer is given in the following discussion.

\begin{widetext}

\begin{figure}[H]
\selectlanguage{english}%
\begin{centering}
\includegraphics[width=0.45\columnwidth]{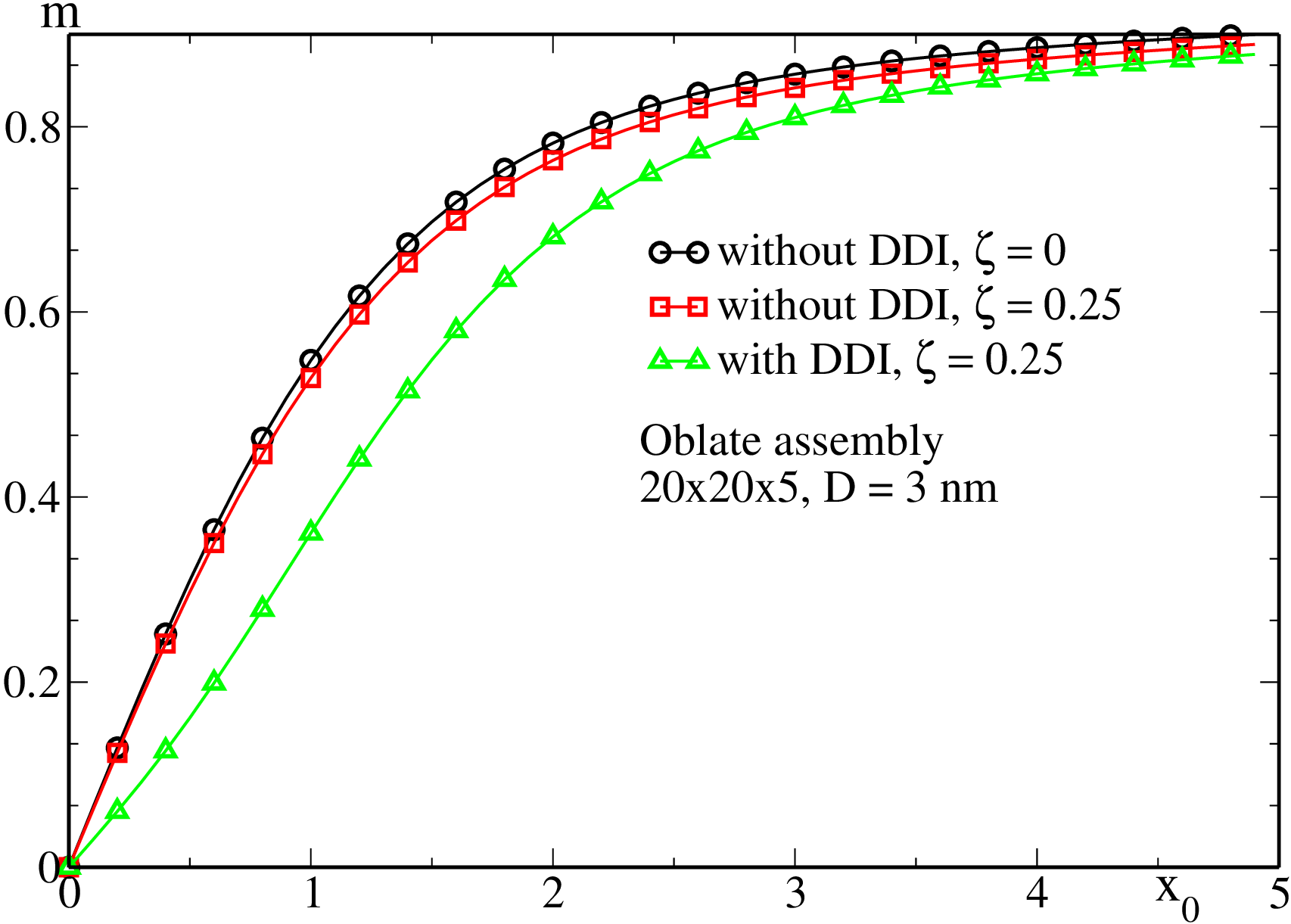}\includegraphics[width=0.45\columnwidth]{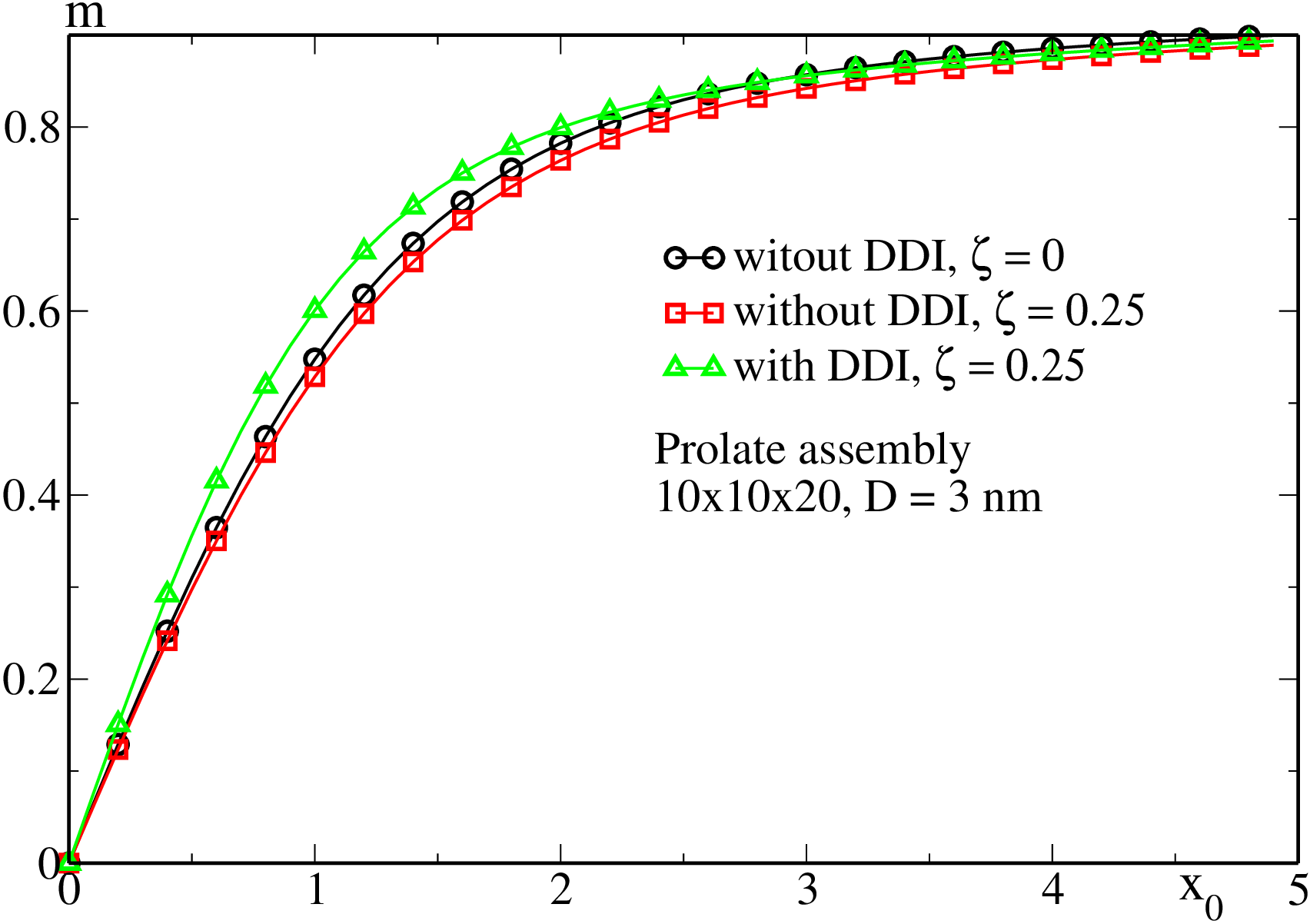} 
\par\end{centering}

\selectlanguage{american}%
\centering{}\caption{(color online) \label{fig:EOSPvsDDI}\foreignlanguage{english}{Left:
magnetization as a function of the (dimensionless) field $x_{0}=\frac{\mu_{B}H}{k_{B}T}$
for an oblate sample ($20\times20\times5$). Right: same for a prolate
sample ($10\times10\times20$). }\foreignlanguage{french}{Here, for
both panels $\sigma=3.275$ and when the dipolar interaction is switched
on $\xi\simeq0.18$ corresponding to a volume concentration $C_{{\rm v}}=0.24\%$.}}
\end{figure}

\end{widetext}

Plots of the magnetization, which take into account both surface effects
and DDI, are shown in Fig. \ref{fig:EOSPvsDDI} as a function of the
(dimensionless) field $x$ for an oblate sample with $N_{x}\times N_{y}\times N_{z}=20\times20\times5$
and a prolate sample with $10\times10\times20$, with the corresponding
values of $\mathcal{C}^{(0,0)}\simeq-4.0856$ and $1.7293$, respectively.
For larger systems, with the same aspect ratios, one obtains, according
to Ref. \textbackslash{}onlinecite\{AharoniJAP1998\}, $\mathcal{C}^{\left(0,0\right)}\simeq-3.9868$
and $1.69662$. For a given sign of $\zeta$, the figure shows the
role of the sample's geometry: whilst the intrinsic surface effects
are always demagnetizing in the present case with $\zeta>0$, the
DDI can either contribute positively (prolate sample) or negatively
(oblate sample) to the magnetization. 

\selectlanguage{american}%

\selectlanguage{english}%
For maghemite nanoparticles it was observed that for dilute assemblies
of very small particles (of $3\,{\rm nm}$ in diameter) the magnetization
curves $m\left(H\right)$ at different temperatures showed a kind
of a ``fanning out'' as the temperature drops below some value (circa
$100\,$ K), see Fig. 3 of Ref. \onlinecite{troetal00jmmm}. The
magnetization enhancement at low temperature strongly depends on the
particles size and was thus attributed to surface effects on account
of various experiments and numerical simulations.\cite{kacetal00jmmm}
However, as the concentration increases this fanning vanishes thus
recovering the usual magnetization curves $m\left(H\right)$ with
regular spacing at different temperatures, see Fig. 5 of Ref. \onlinecite{troetal00jmmm}.
This means that DDI, which become stronger in more concentrated assemblies,
seem to ``screen out'' surface effects and thus to compensate for
them {[}see discussion in Ref. \onlinecite{kachkachi06neim}{]}. In
fact, this compensation is only partial because the magnetization
still does not saturate at the highest field available, as this can
be seen in Fig. 5 of Ref. \onlinecite{troetal00jmmm}. In the light
of the present calculations, the results of Ref. \onlinecite{troetal00jmmm}
{[}see also Ref. \onlinecite{kacetal00epjb}{]} seem to imply that
DDI have an opposite effect to that of surface anisotropy, represented
here by the contribution in $\zeta$. More precisely, according to
Eqs. (\ref{eq:MagEOSPvsDDI1}, \ref{eq:MagEOSPvsDDI2}) this implies
that the product $\tilde{\xi}\zeta$ is negative, and considering
the fact that $\tilde{\xi}<0$ for oblate samples, the results Ref.
\onlinecite{troetal00jmmm} would imply that $\zeta>0$. Note, however,
that the comparison of the present calculations and the experimental
results \cite{troetal00jmmm} is done with a little daring because
the measurements of Ref. \onlinecite{troetal00jmmm} were done on
thin disc-shaped (hence oblate) samples with the magnetic field applied
along the sample plane. In addition, the particles effective anisotropy
easy axes are randomly distributed.

As discussed in the introduction, this result may help in optimizing
the physical parameters (size, shape, concentration, etc.) in view
of further fundamental investigations, \emph{e.g.} of the dynamical
properties and the effect of surface anisotropy.

\section{\label{sec:Results-and-discussions}Discussion}

Now we present and discuss plots of the magnetization for assemblies
with varying parameters. First, we discuss the effect of DDI alone,
without surface effects ($\xi\ne0$, $\zeta=0$). Next, we comment
on the plots of the EOSP calculations of Section \ref{sec:EOSP-vs-DDI}
showing the combined effects of DDI and surface anisotropy ($\xi\ne0$,
$\zeta\neq0$).

\subsection{\label{sub:Effect-of-dipolar}Effect of dipolar interactions on the
magnetization ($\zeta=0$ \& $\xi\neq0$)}

Here we restrict ourselves to the study of the DDI effect \foreignlanguage{french}{$(\xi\ne0)$}
on the magnetization of an assembly of monodisperse magnetic nanoparticles,
ignoring for the time being the effect of surface anisotropy ($\zeta=0$).
The results of magnetization curves at different temperatures for
assemblies with different concentrations (increasing downwards) and
particle diameters (increasing from left to right) are shown in Fig.
\ref{fig:Aimantation-d'une-assembl=0000E9e}.

\selectlanguage{american}%
\begin{widetext}

\selectlanguage{french}%
\begin{figure}[H]
\begin{centering}
\begin{tabular}{ccc}
\includegraphics[scale=0.2]{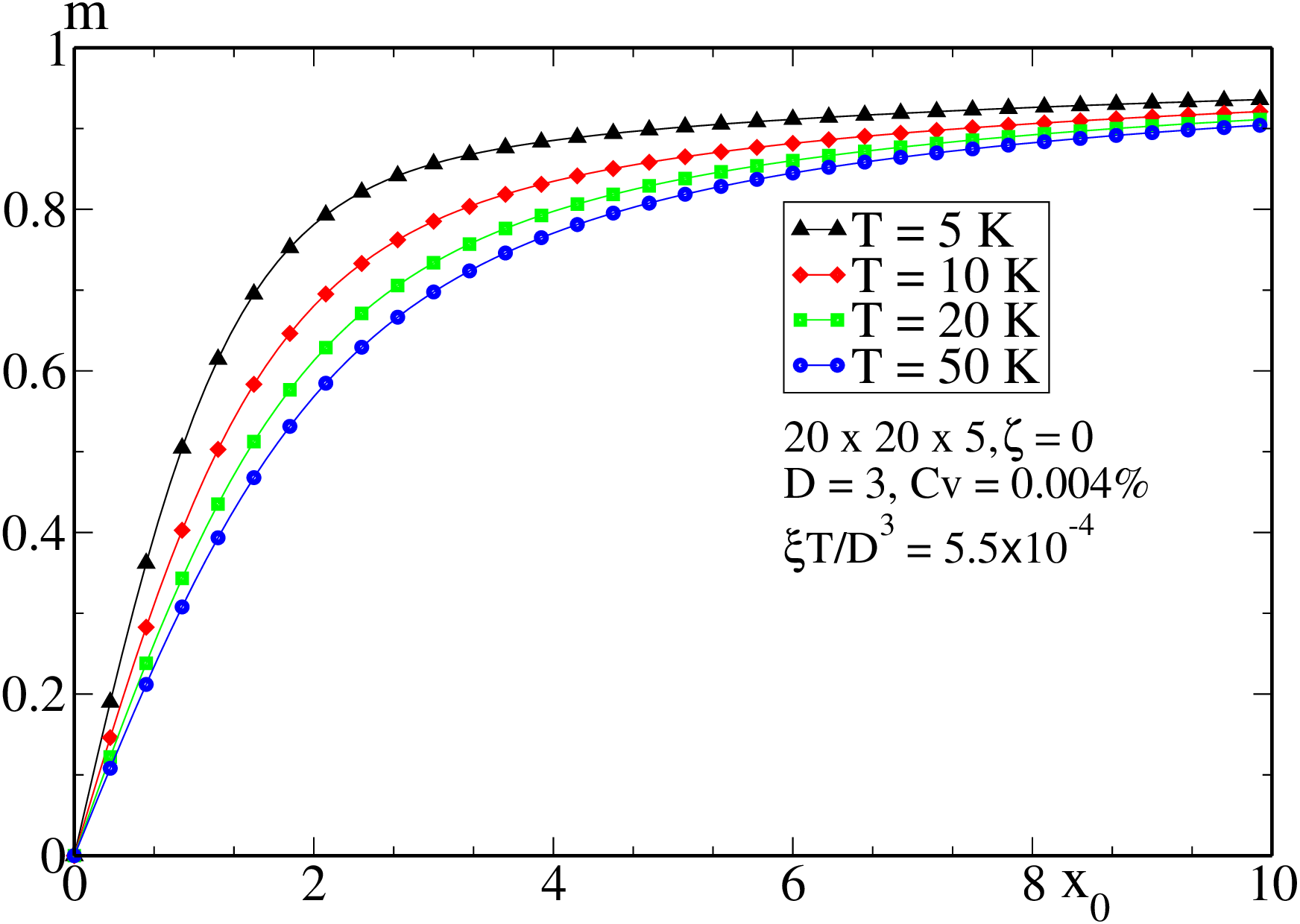} & \includegraphics[scale=0.2]{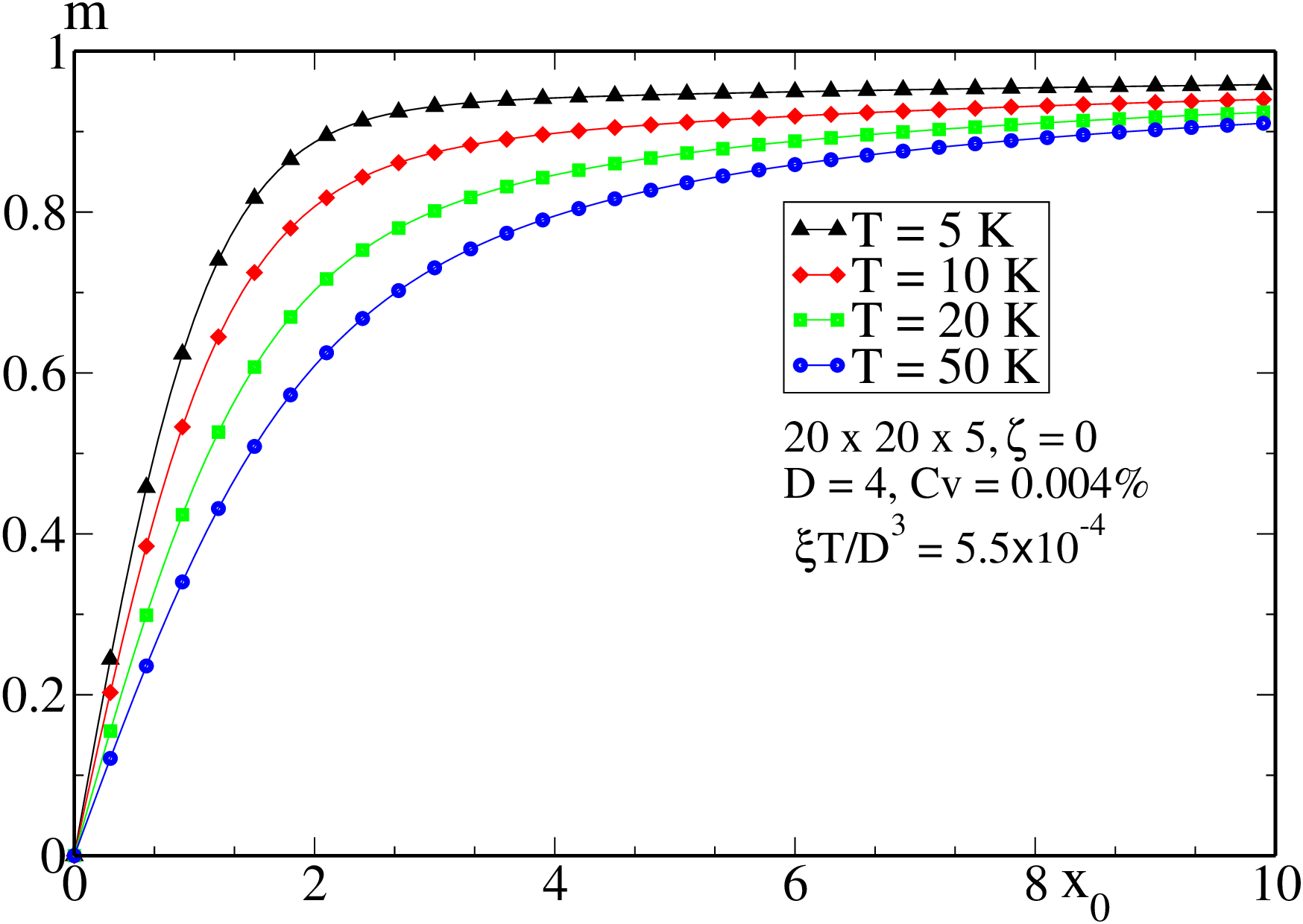} & \includegraphics[scale=0.2]{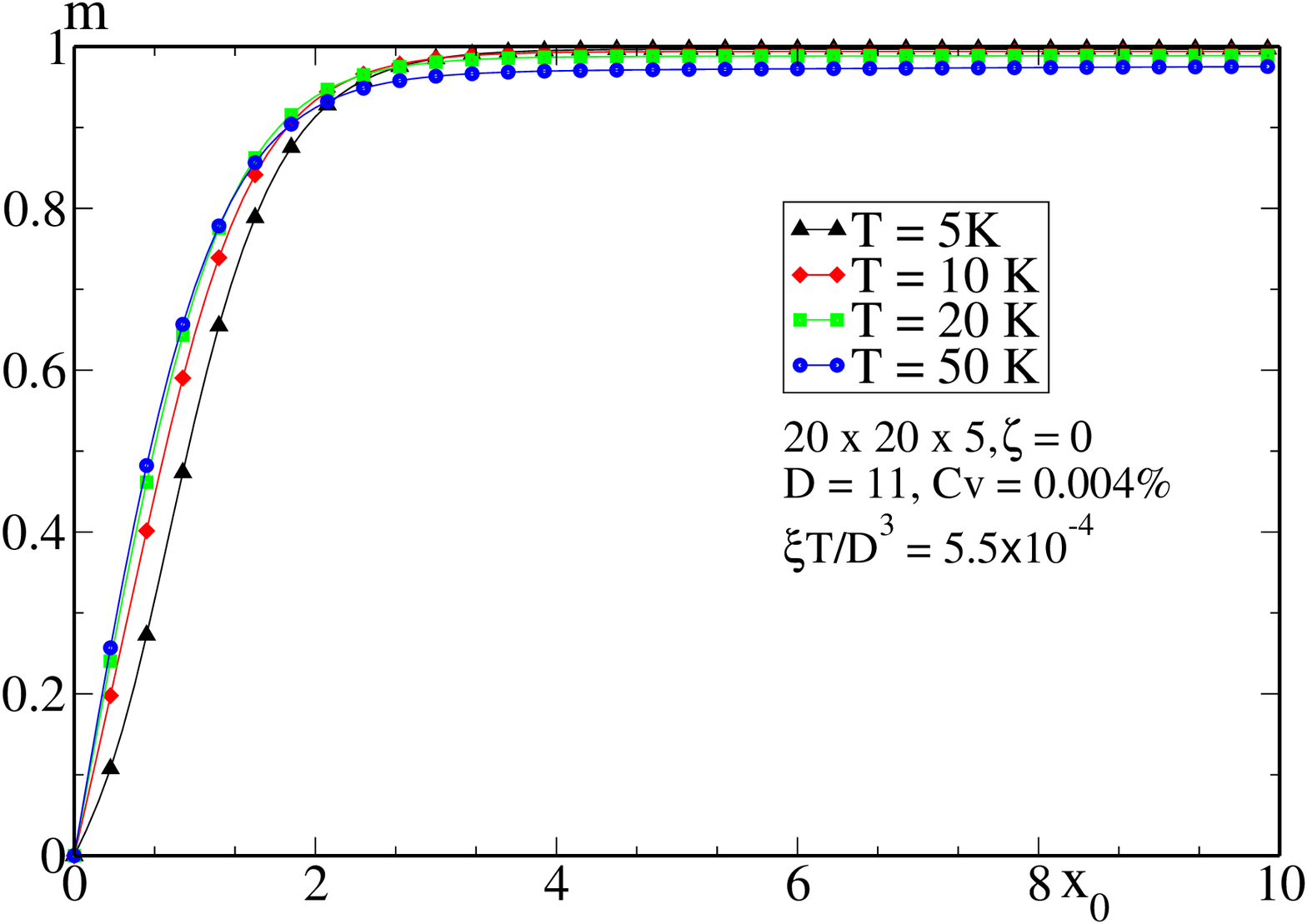}\tabularnewline
\includegraphics[scale=0.2]{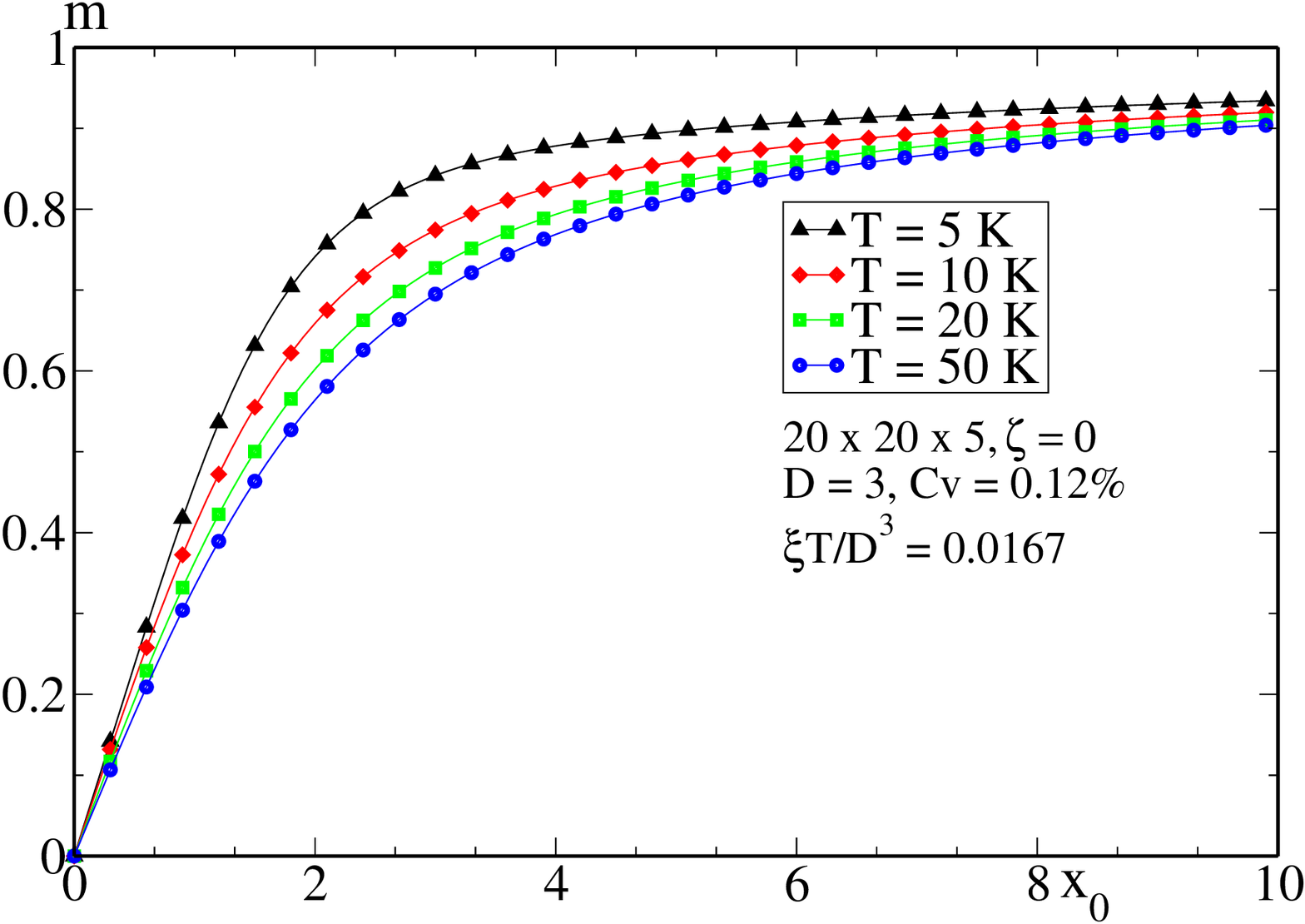} & \includegraphics[scale=0.2]{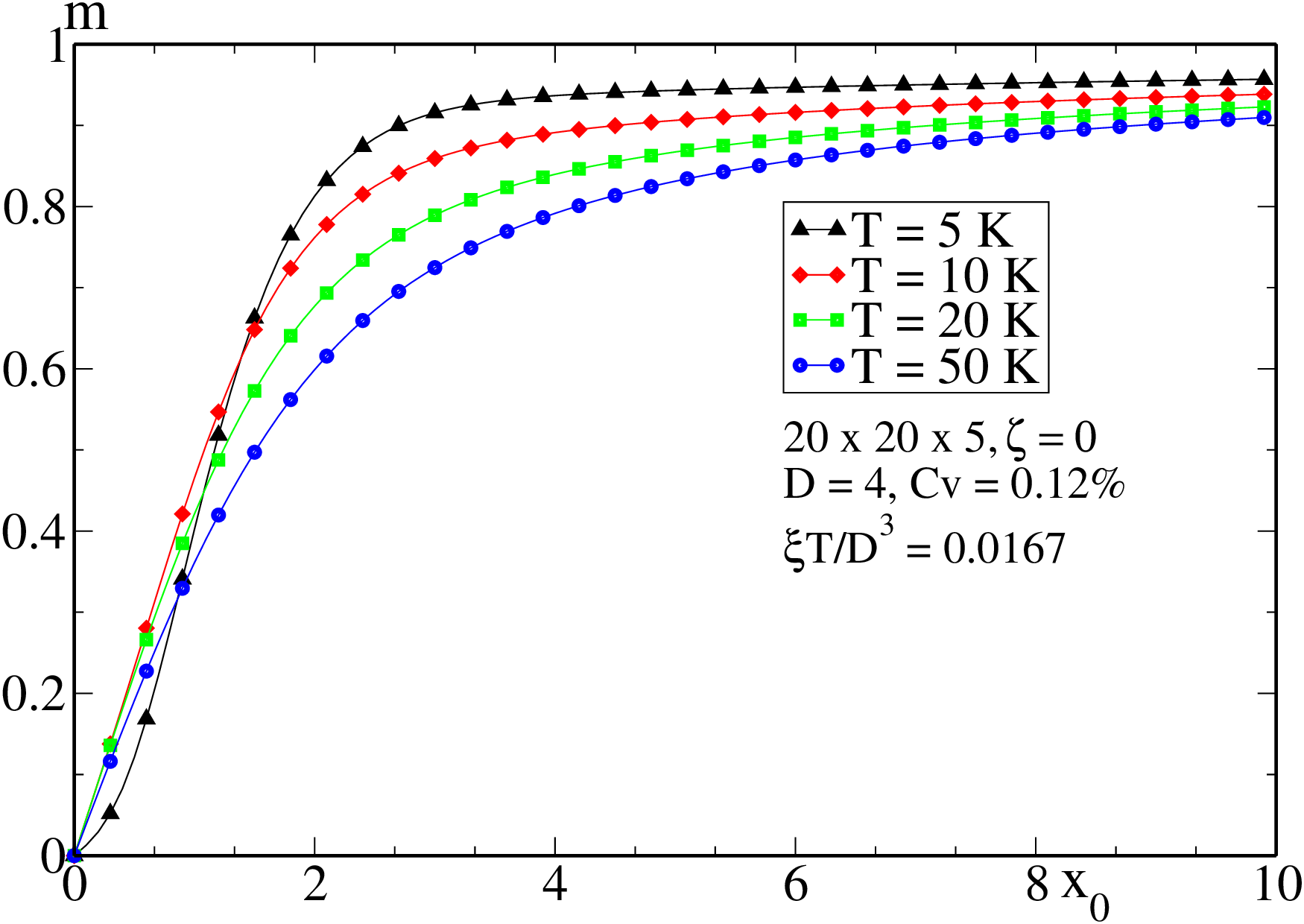} & \selectlanguage{english}%
\selectlanguage{french}%
\tabularnewline
\includegraphics[scale=0.2]{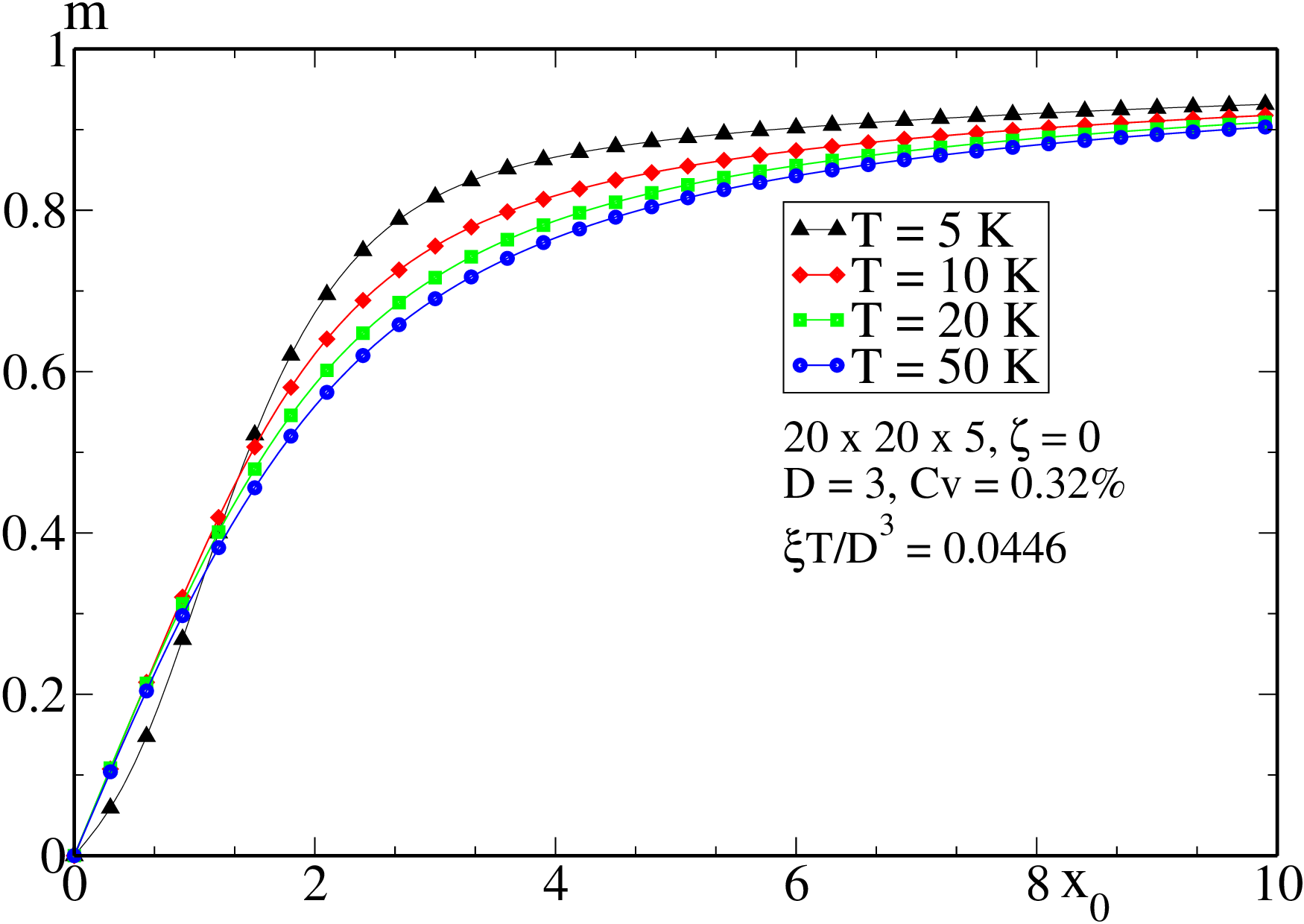} & \selectlanguage{english}%
\selectlanguage{french}%
 & \selectlanguage{english}%
\selectlanguage{french}%
\tabularnewline
\end{tabular}
\par\end{centering}

\centering{}\caption{(color online) \label{fig:Aimantation-d'une-assembl=0000E9e}\foreignlanguage{english}{\textcolor{red}{{}
}Magnetization curves of an oblate sample ($20\times20\times5$) as
a function of the (dimensionless) field $x_{0}=\frac{\mu_{B}H}{k_{B}T}$,
for different temperatures $T$, different particle diameters $D$
(in nm) and different volume concentrations $C_{{\rm v}}$, without
surface effects ($\zeta=0\ \mbox{ and}\ \xi\neq0$). Since the reduced
(uniaxial) anisotropy $\sigma$ depends on both the temperature and
the particle diameter, the calculations presented in these panels
are done for a constant value of $\frac{\sigma T}{D^{3}}\simeq0.6066\ {\rm K.nm^{-3}}$.
Similarly, for each line the dipolar parameter $\frac{\xi T}{D^{3}}$
is expressed in the same units. The upper panel with a very low concentration
($C_{{\rm v}}=0.004\%$) corresponds to an inter-particle distance
of $a\simeq15.3D$, while the highest concentration in the lower panel
($C_{{\rm v}}=0.32\%$) leads to $a\simeq3.5D$.}}
\end{figure}

\selectlanguage{american}%
\end{widetext}

\selectlanguage{english}%
The upper right panel displays the magnetization curves of an assembly
with a volume concentration $C_{{\rm v}}=0.004\%$ and a particle
diameter $D=11\ {\rm nm}$. In this case, we see that the saturation
is obviously easier at lower temperatures. For a given concentration,
upon increasing the diameter of the particles (thus moving to the
right within the same row), we see that the magnetization curves saturate
more quickly (in lower fields) since for larger particles the Zeeman
energy ($\mathcal{E}_{\mathrm{Zeeman}}\propto{\bf m}\cdot{\bf H}$
with $m\propto n_{i}$) is larger and thence the magnetizing effect
of the external field is bigger. On the other hand, for a given particle
diameter, an increase of the particle concentration $C_{{\rm v}}$\emph{
}(going downwards within the same column) reduces the inter-particle
distance $a$ and thus increases the DDI parameter $\tilde{\xi}$.
For the oblate sample considered here the DDI tend to maintain the
magnetic moments in the $xy$ plane and thus oppose the effect of
the external magnetic field. This competition is reflected in the
magnetization curves by the appearance of an inflection point which
is well marked for the black curves in Fig. \ref{fig:Aimantation-d'une-assembl=0000E9e},
\emph{i.e. }for $T=5\ {\rm K}$. This feature becomes obvious by looking
at the susceptibility, namely $dm/dx_{0}$, as displayed in Fig. \ref{fig:Aimantation_et_derive}. 

The results are in agreement with the low field expansion (\ref{eq:MagEOSPvsDDI1})
which reads (for $\zeta=0$) 

\begin{align*}
m & \simeq\left(1-\frac{1}{\sigma}\right)x-\left(1-\frac{2}{\sigma}\right)\frac{x^{3}}{3}\\
 & +\tilde{\xi}\left[\left(1-\frac{2}{\sigma}\right)x-4\left(1-\frac{3}{\sigma}\right)\frac{x^{3}}{3}\right].
\end{align*}

Indeed, at low fields the first line (corresponding to free particles)
in this expression goes above the one between the square brackets
when both plotted against the field. As such, as the concentration
increases, $\tilde{\xi}$ increases and the effect of DDI tends to
depress the magnetization. We may use different terms to interpret
the appearance of the inflection point for constant volume concentration
$C_{{\rm v}}$ when $D$ is increased, as it is the case when comparing
the $T=5\ {\rm K}$ curves of the upper panels in Fig. \ref{fig:Aimantation-d'une-assembl=0000E9e}.
Going to larger particles increases $\tilde{\xi}$ which leads to
a sign change of the second derivative of $m$ with respect to the
field as it can be inferred from the expression above. This well marked
feature in the magnetization curve can be viewed as the signature
of dipolar interaction in textured oblate assemblies. In the case
of a prolate sample, the DDI induce an anisotropy that adds up to
the magneto-crystalline anisotropy that is intrinsic to the particles.
Consequently, there is no competition between DDI and the latter.

\subsection{\label{sub:Effective-model-EOSP}Effective model EOSP ($\zeta\neq0$
\& $\xi\neq0$)}

In order to take account of surface effects we include effective anisotropy
contribution ($\zeta\ne0$) according to the EOSP model. We will show
here that depending on the sign of $\zeta$, we can have concomitant
or competing effects between surface and dipolar contributions.

\begin{figure}[H]
\selectlanguage{english}%
\begin{centering}
\includegraphics[width=1\columnwidth]{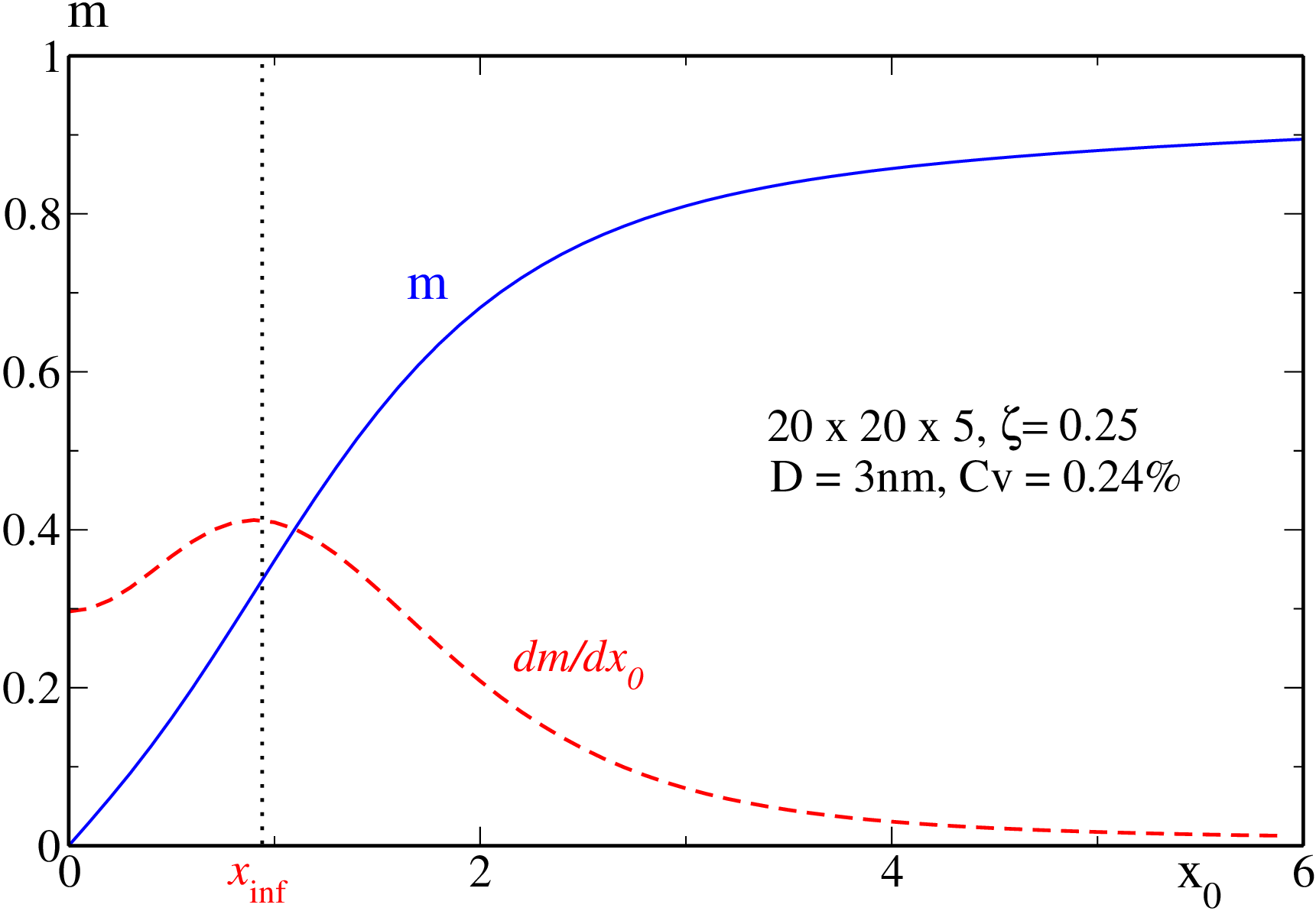}
\par\end{centering}

\selectlanguage{french}%
\centering{}\caption{\label{fig:Aimantation_et_derive}\foreignlanguage{english}{(color
online) Magnetization and its derivative as functions of the (dimensionless)
field $x_{0}=\frac{\mu_{B}H}{k_{B}T}$ and its derivative for an oblate
assembly at $T=5K$. }}
\end{figure}

We have assumed DDI to be relatively weak so that we can use perturbation
theory to derive an analytical expression for the magnetization. As
such, we have to content ourselves with a small effect of DDI. In
order to understand their interplay with surface effects, we have
studied the behavior of a physical quantity for which their effect
is more explicit. As mentioned earlier it happens that the magnetization
curves present an inflection point that can be more clearly appreciated
by examining the derivative of these curves. This derivative turns
out to exhibit a maximum at some reduced field $x_{\mathrm{inf}}$.
We show an example of this in Fig. \ref{fig:Aimantation_et_derive},
where the inflection point appears at $x_{\mathrm{inf}}\approx0.8$
for the given parameters.

\selectlanguage{french}%

\begin{figure}[H]
\selectlanguage{english}%
\begin{centering}
\includegraphics[scale=0.3]{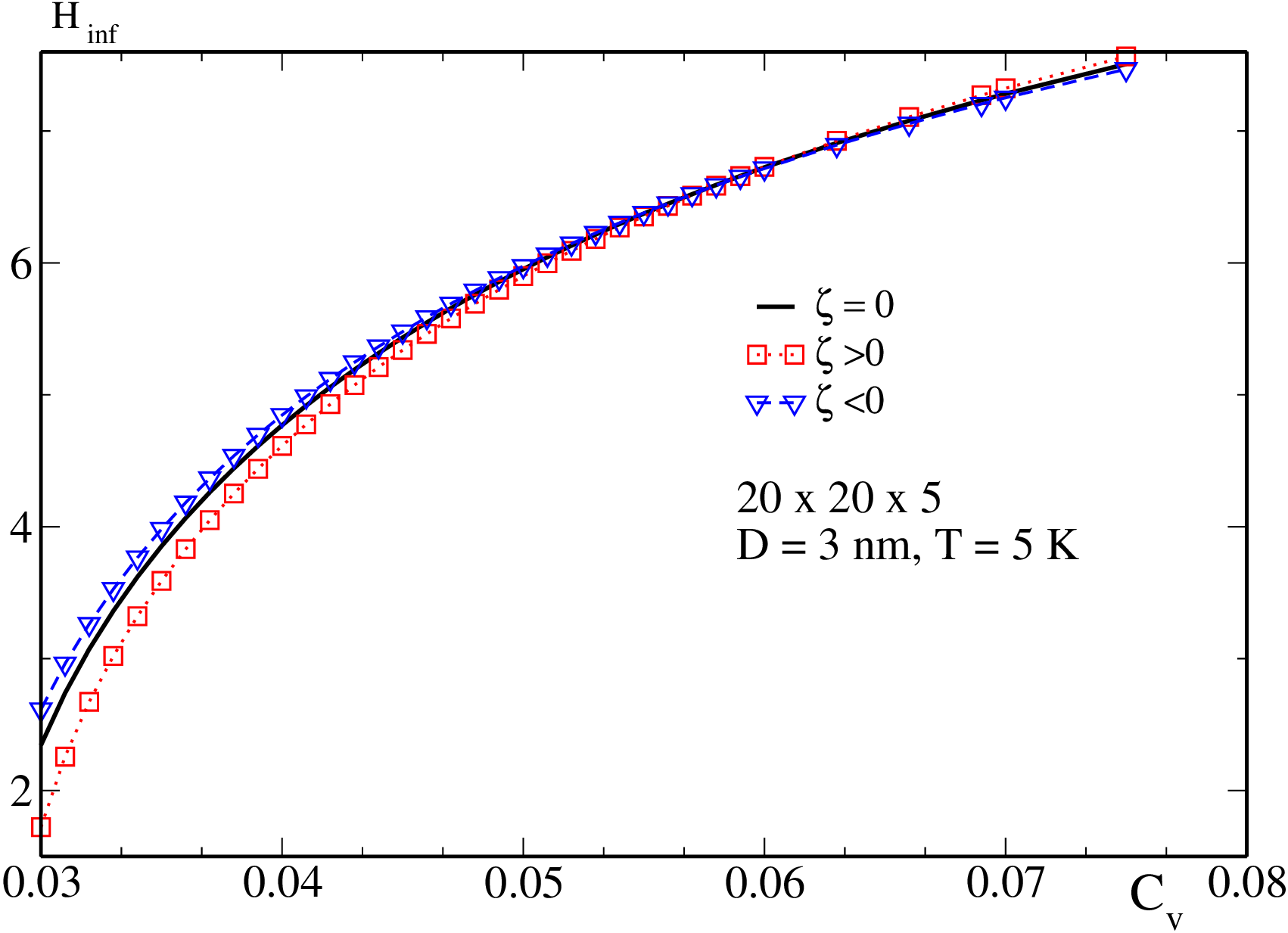}
\par\end{centering}

\selectlanguage{french}%
\centering{}\caption{\label{fig:Hmax_derivee}\foreignlanguage{english}{ (color online)
Field $H_{\mathrm{inf}}$ as a function of $C_{{\rm v}}$ for a monodisperse
oblate $(20\times20\times5$) assembly with $D=3\ \mathrm{nm}$ and
$T=5\, K$. }}
\end{figure}

\selectlanguage{english}%
We then extract $x_{\mathrm{inf}}$ (or $H_{\mathrm{inf}}$) and plot
it as a function of the concentration, or equivalently the DDI coefficient
$\tilde{\xi}$. This is shown in Fig\foreignlanguage{french}{. \ref{fig:Hmax_derivee}}
that applies to a monodisperse assembly of nanoparticles with diameter
$D=3\ {\rm nm}$ at $T=5\ {\rm K}$, for a sample of size \foreignlanguage{french}{$\left(20\times20\times5\right)$}.
There are three curves : one for $\zeta=0$, for which surface effects
are dropped, one for $\zeta<0$, and the other for $\zeta>0$ for
which surface effects play opposite roles. The change of sign of $\zeta$
can be achieved experimentally owing to the fact that, depending on
the material chosen for the particles and their size and shape, the
effective anisotropy can change sign. 

In Fig.\foreignlanguage{french}{ \ref{fig:Hmax_derivee}} we first
see that $H_{\mathrm{inf}}$ increases with $C_{{\rm v}}$ as expected
for all cases since then DDI become stronger and the competition with
the external magnetic field becomes more pronounced. Let us compare
the curve $\zeta<0$ to that with $\zeta=0$, keeping in mind our
discussion of the effects of a cubic anisotropy. The various contributions
to the energy are : i) the uniaxial anisotropy with an easy axis along
the $z$ axis, ii) the cubic anisotropy for which the easy axes are
along $x,y,$ and $z$ for $\zeta<0$, iii) the external magnetic
field along $z$, and iv) DDI which tend to place the magnetic moments
in the $xy$ plane for an oblate sample ($\zeta<0$).

\selectlanguage{french}%

\selectlanguage{english}%
For low concentrations, the uniaxial and cubic anisotropies are the
most dominant contributions to the energy. The three directions $x,$
y, and $z$ are degenerate with regard to the cubic anisotropy, but
since the $z$ axis is favored by the uniaxial anisotropy, this direction
is selected. For this reason the field $H_{\mathrm{inf}}$ is larger
if $\zeta<0$. On the opposite, for high concentrations the DDI contribution
is predominant and its effect is concomitant with that of the cubic
anisotropy. Again, the directions $x,$ $y$, and $z$ are degenerate
with regard to the cubic anisotropy, but this time the $xy$ plan
is favored by DDI. It is therefore more difficult to drive the magnetization
out of the $xy$ plane and this explains why $H_{\mathrm{inf}}$ becomes
lower than for the $\zeta=0$ case above some value of $C_{{\rm v}}$. 

For a more systematic study, we compared an oblate ($20\times20\times5$)
with a prolate ($10\times10\times20$) sample. The results are shown
in Fig. \ref{fig:Competition_entre_ES_DDI_Aimantation}, where the
magnetization is plotted as a function of the field $x$. We consider
different concentrations $C_{{\rm v}}$, and different situations
with respect to surface effects by choosing either $\zeta>0$ or $\zeta<0$.
As a reference, we also plot the curve corresponding to free particles
($C_{{\rm v}}\sim0$) with and without surface effects ($\zeta=0$).

\selectlanguage{american}%
\begin{widetext}

\selectlanguage{french}%
\begin{figure}[H]
\begin{centering}
\begin{tabular}{cc}
\includegraphics[width=0.43\columnwidth]{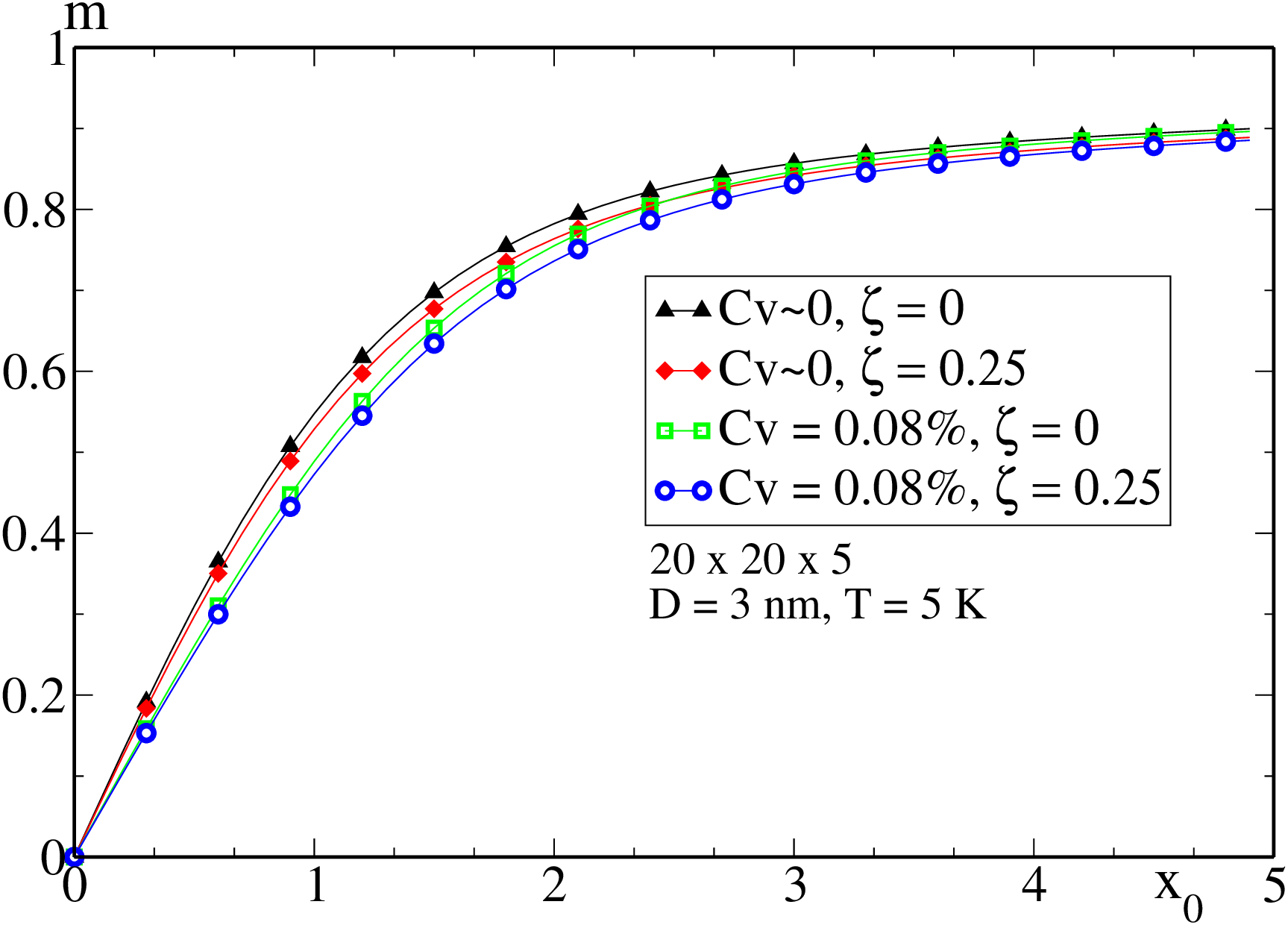} & \includegraphics[width=0.43\columnwidth]{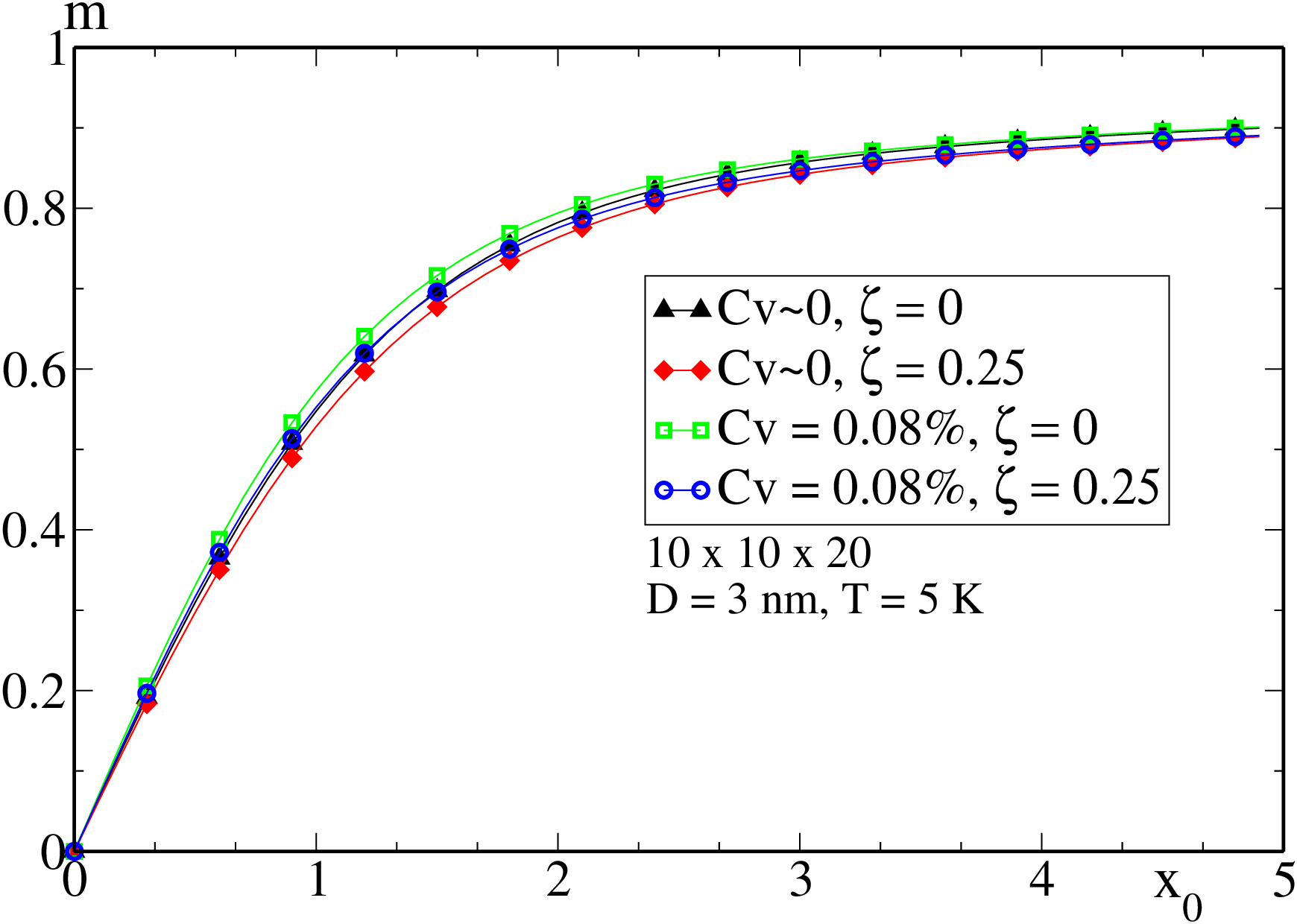}\tabularnewline
\includegraphics[width=0.43\columnwidth]{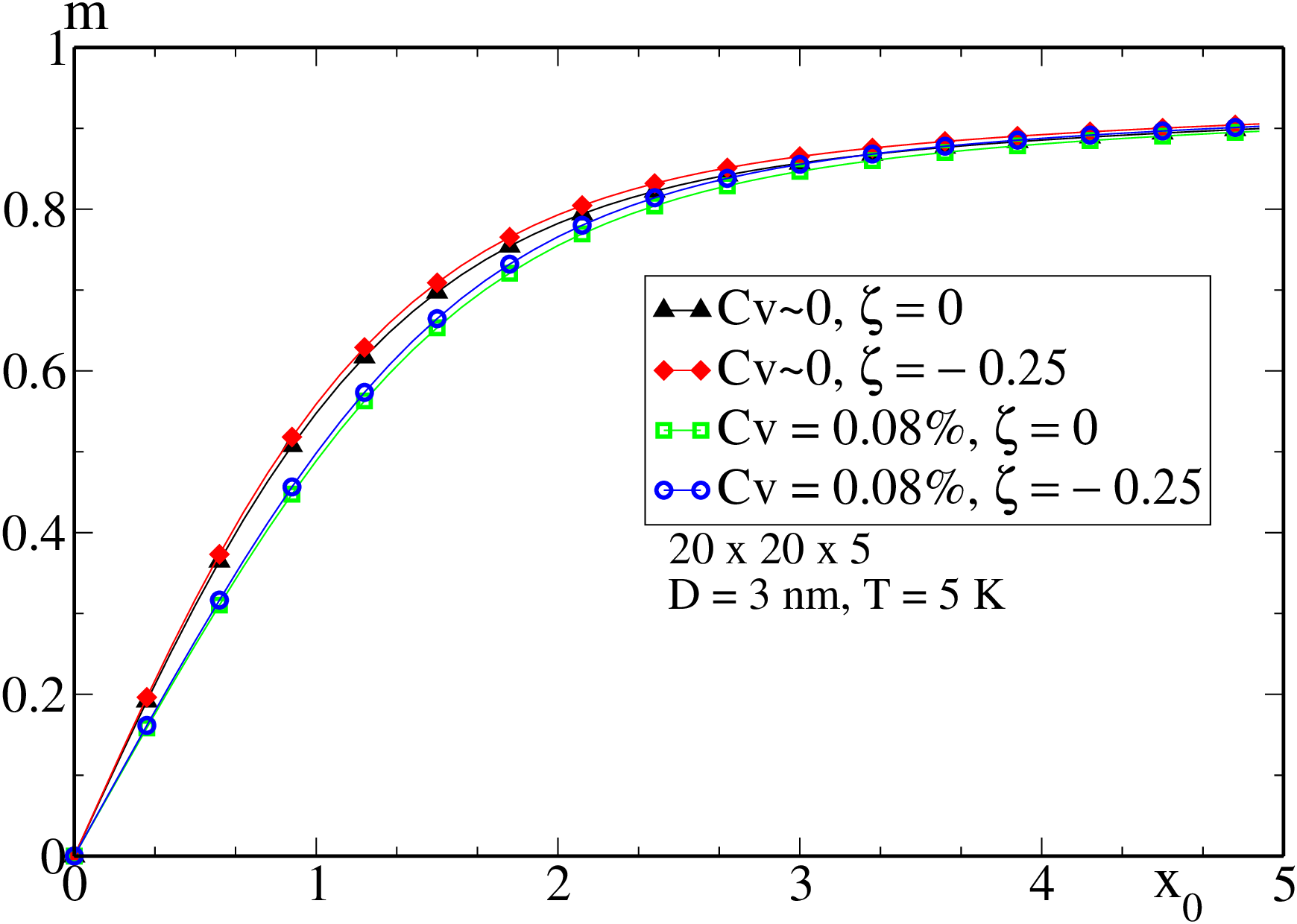} & \includegraphics[width=0.43\columnwidth]{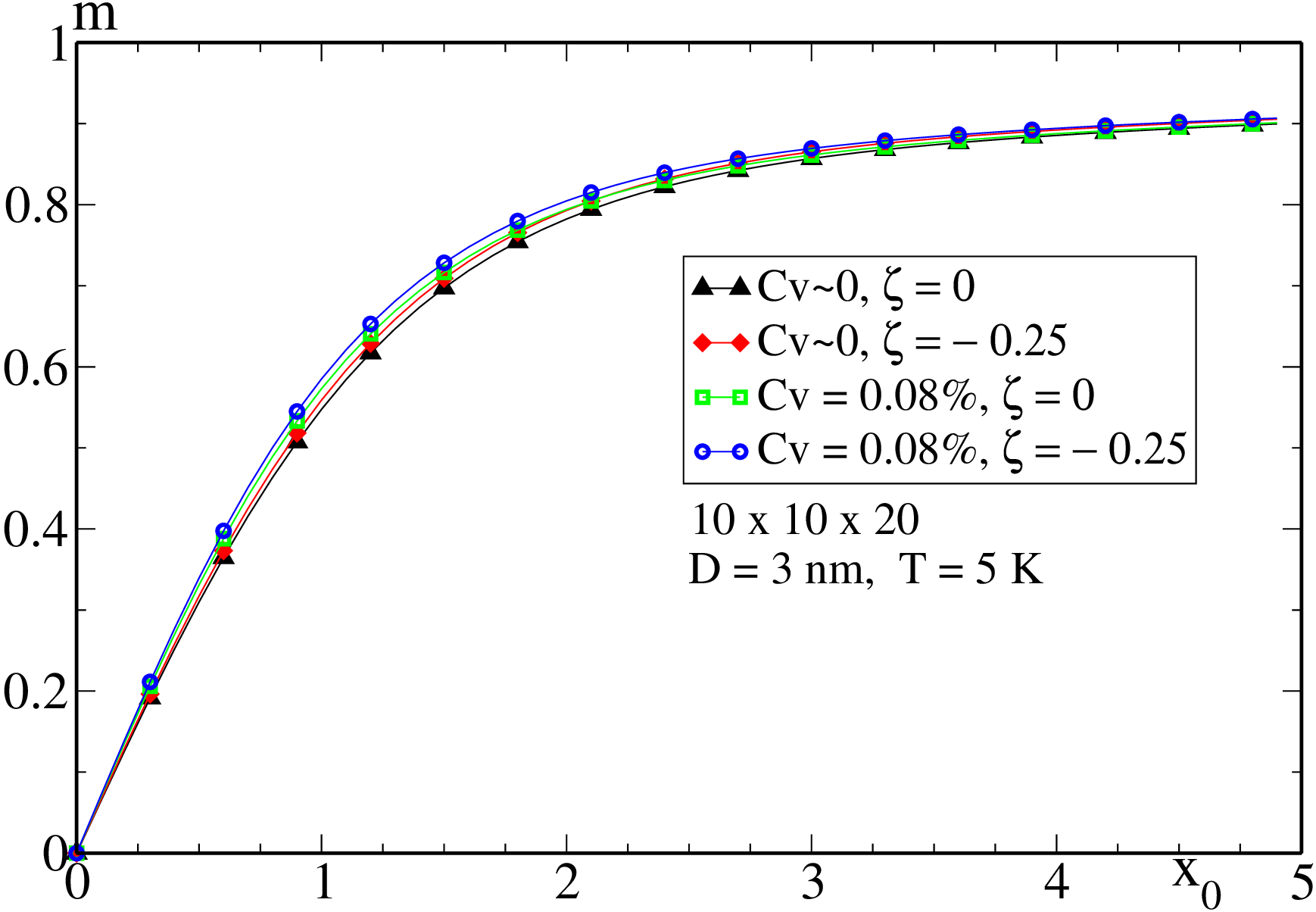}\tabularnewline
\end{tabular}
\par\end{centering}

\centering{}\caption{\label{fig:Competition_entre_ES_DDI_Aimantation}\foreignlanguage{english}{(color
online) Magnetization as a function of the (dimensionless) field ${\color{magenta}{\normalcolor x}_{{\normalcolor 0}}}$
for assemblies of monodisperse particles of diameter D = $3\ \mathrm{nm}$
at (constant) temperature $T=5K$. Left column: oblate $(20\times20\times5)$
sample. Right column: prolate $(10\times10\times20)$ sample.}}
\end{figure}

\selectlanguage{american}%
\end{widetext}

\selectlanguage{english}%
For the oblate sample $\left(20\times20\times5\right)$ the easy plane
for DDI is the $xy$ plane. The upper left graph of Fig. \ref{fig:Competition_entre_ES_DDI_Aimantation}
corresponds to $\zeta>0$ for which the cubic anisotropy easy axes
are along the main diagonals of the cube. DDI and surface effects
cooperate to suppress the magnetization leading to a magnetization
curve that remains below the reference curve. The graph at the bottom
left of Fig. \ref{fig:Competition_entre_ES_DDI_Aimantation} represents
the case with $\zeta<0$. We observe in this case a competition between
DDI and surface effects. 

For the prolate sample $\left(10\times10\times20\right)$, the opposite
effect is observed. Indeed, this time the $z$ axis is favored by
DDI. If $\zeta<0$, as shown in the bottom right of Fig. \ref{fig:Competition_entre_ES_DDI_Aimantation},
the $z$ axis is selected by both DDI and surface anisotropy. Hence,
the magnetization curves fall above the reference curve and thus the
magnetization of the assembly is enhanced.

\section{\label{sec:Conclusion}Conclusion and Outlook}

In this work we have examined in detail the competition between intrinsic
effects and collective behavior in an assembly of (weakly) interacting
nanoparticles. In order to account for surface effects we have adopted
the effective one-spin problem that represents a nanoparticle as a
macroscopic magnetic moment evolving in an effective energy potential
which is a polynomial in the components of the particle's net magnetic
moment. The coefficients of this potential change in sign and magnitude
with the underlying lattice, the size, and the shape of the particles. 

This, in conjunction with the use of perturbation theory, has allowed
us to derive sensible analytical expressions for the magnetization
of a nanoparticle within the assembly that include the applied magnetic
field, the core and surface anisotropy, and the dipole-dipole interactions. 

Given the fact that i) the DDI contribution changes sign according
to the shape of the assembly and ii) the surface-anisotropy contribution
also changes sign and magnitude upon changing the underlying lattice,
the size, and the shape of the particles, it is possible to design
new samples where various situations can be encountered with either
competing or concomitant effects. In Ref. \onlinecite{dejardinetal08jpd}
it was shown that as the surface anisotropy increases, the magnetization
relaxation rate evolves through a bell-like shaped curve, which means
that there is a range of physical parameters (lattice, size, shape,
surface morphology) for which the relaxation rate is maximal or the
relaxation time is the shortest. This implies that for such parameters
the particle magnetization is least stable against thermal fluctuations.
Now, nanoparticles with physical parameters outside this range may
be organized or assembled in such a way that the dipole-dipole interactions
acquire a concomitant effect that eventually leads to an assembly
with higher and more stable magnetization. This is one of the main
features sought for future applications $e.g.$ in the area of magnetic
recording.

From the standpoint of fundamental research, we hope that this work
sheds a new light on the behavior of such a complex system as an interacting
assembly of nanoparticles whose intrinsic features are accounted for
to some extent. Indeed, the analytical expressions for the magnetization,
established in various typical realistic regimes, are a handy tool
for further investigations of the effects of the various physical
parameters involved. Likewise, they will be quite useful in accompanying
the near future numerical simulations using the Monte Carlo technique
that we have planned and in which we intend to test this competition
between surface anisotropy and dipolar interactions. More precisely,
we consider an assembly of a small number of nanoclusters modeled
as many-spin systems. The idea behind this work is, in addition to
a comparison with the analytical work developed here, to explore the
regime of stronger dipolar interactions and to assess the limit of
validity of the results presented here. 

The effects of surface anisotropy on the dynamics of a nanoparticle
assembly is one of the still appealing topics with a plethora of physical
phenomena awaiting more thorough investigation. Accordingly, we are
planning new equilibrium and dynamical measurements on newly produced
chains of iron nanoparticles with controlled size and separation.\cite{toulemonetal11ChemComm,paulyetal12JMatChem}
One of the issues we would like to investigate is related with the
effect of particle separation on the low- and high-frequency dynamics
of the nanoparticles. The results of the present work will be of precious
help in this endeavor.
\begin{acknowledgments}
We would like to thank Denis Ledue for fruitful discussions of dipolar
interaction in assemblies of magnetic nanoparticles. O. Iglesias acknowledges
funding by the Spanish MINECO (MAT2009-0667 and MAT2012-33037), Catalan
DURSI (2009SGR856) and European Union FEDER funds (Una manera de hacer
Europa). He also acknowledges CESCA and CEPBA under coordination of
C4 for supercomputer facilities.
\end{acknowledgments}
\appendix

\section{Magnetization and high-order correlations}

We give here some details of the calculation of the equilibrium magnetization
for the EOSP model without DDI. Using the Legendre polynomials, $p_{n}(z)$
{[}$p_{1}(z)=z,\, p_{2}(z)=\frac{1}{2}(3z^{2}-1),\ p_{4}(z)=\frac{1}{8}\left(35x^{4}-30x^{2}+3\right)\ldots${]}
we define the anisotropy-weighted averages 
\begin{equation}
\mathcal{C}_{n}\left(\sigma,x\right)=\left\langle p_{n}\left(z\right)\right\rangle _{0}=\int_{-1}^{1}dz\,\mathcal{P}_{0}\left(z\right)\, p_{n}\left(z\right).\label{eq:Cn_fonction}
\end{equation}
Let us also define the equivalent averages at zero field, \emph{i.e.}
$S_{n}\left(\sigma\right)\equiv\mathcal{C}_{n}\left(\sigma,0\right).$
One finds 
\begin{eqnarray*}
 &  & \left\langle s_{z}\right\rangle _{0}=\mathcal{C}_{1},\quad\left\langle s_{z}^{2}\right\rangle _{0}=\dfrac{1}{3}\left(1+2\mathcal{C}_{2}\right),\\
 &  & \left\langle s_{x,y}\right\rangle _{0}=0,\quad\left\langle s_{x,y}^{2}\right\rangle _{0}=\dfrac{1}{3}\left(1-\mathcal{C}_{2}\right)
\end{eqnarray*}
 with
\begin{eqnarray}
\mathcal{C}_{1} & = & \dfrac{e^{\sigma}}{\sigma Z_{\parallel}^{\left(0\right)}}\sinh x-h=m^{\left(0\right)},\label{eq:12Moments}\\
\mathcal{C}_{2} & = & \dfrac{3}{2}\left[\dfrac{e^{\sigma}}{\sigma Z_{\parallel}^{\left(0\right)}}\left(\cosh x-h\sinh x\right)+h^{2}-\dfrac{1}{2\sigma}\right]-\dfrac{1}{2}.\nonumber \\
\mathcal{C}_{4} & = & \frac{3}{2}\frac{1}{Z_{\parallel}^{(0)}\sigma}\left[(2\exp\sigma-\mathcal{C}_{2}-\frac{1}{\sigma})\cosh x\right]\nonumber \\
 &  & +\frac{3}{4}\frac{x\sinh x}{Z_{\parallel}^{(0)}\sigma}\left[\mathcal{C}_{2}+\frac{1}{\sigma}\right]+\frac{3}{2}\left[\frac{x^{2}}{3\sigma^{3}}+\frac{1}{2\sigma^{2}}\right]\nonumber 
\end{eqnarray}

Then the (reduced) equilibrium susceptibility tensor is defined as
\begin{eqnarray*}
 &  & \chi_{\alpha\beta}=\left\langle s_{\alpha}s_{\beta}\right\rangle _{0}-\left\langle s_{\alpha}\right\rangle _{0}\left\langle s_{\beta}\right\rangle _{0},\\
\\
 &  & \chi_{\parallel}=\chi_{zz},\quad\chi_{\perp}=\chi_{xx}=\chi_{yy}.
\end{eqnarray*}
Using the definitions given above, the (reduced) static susceptibility
components read 
\begin{equation}
\chi_{\parallel}=\dfrac{1+2S_{2}}{3}-S_{1}^{2},\qquad\chi_{\perp}=\dfrac{1-S_{2}}{3}.\label{eq:StaticXi}
\end{equation}
We then define\foreignlanguage{american}{ }
\begin{eqnarray}
Z_{\parallel}^{\left(n\right)} & \equiv & \int s_{z}^{2n}d\omega^{\left(0\right)}=\frac{\partial^{n}Z_{\parallel}^{\left(0\right)}}{\partial\sigma^{n}}\label{eq:FreePFn}
\end{eqnarray}
and write the first two derivatives of $Z_{\parallel}^{\left(0\right)}$
with respect to $\sigma$, \emph{i.e.} $Z_{\parallel}^{\left(1\right)},Z_{\parallel}^{\left(2\right)}$,
in terms of the averages of Legendre polynomials as follows

\begin{eqnarray}
Z_{\parallel}^{\left(1\right)} & = & \int s_{z}^{2}d\omega^{\left(0\right)}\label{eq:Z0Derivs}\\
 & = & \frac{Z_{\parallel}^{\left(0\right)}}{3}\left(2\mathcal{C}_{2}+1\right),\nonumber \\
Z_{\parallel}^{\left(2\right)} & = & \int s_{z}^{4}d\omega^{\left(0\right)}\nonumber \\
 & = & \frac{Z_{\parallel}^{\left(0\right)}}{35}\left[8\mathcal{C}_{4}+20\mathcal{C}_{2}+7\right].\nonumber 
\end{eqnarray}
$\partial^{2}Z_{0}^{\left(0\right)}/\partial\sigma^{2}$ is simply
the derivative of $\chi_{\parallel}$ above with respect to $\sigma$.
In the absence of the field the averages $\mathcal{C}_{2}$ and $\mathcal{C}_{4}$
become the ``anisotropy functions'' $S_{2}$ and $S_{4}$ which
have the asymptotes \citep{garpal00acp}

\begin{equation}
S_{l}\left(\sigma\right)\simeq\left\{ \begin{array}{lll}
\frac{2^{l/2}\left(l-1\right)!!}{\left(2l+1\right)!!}\sigma^{l/2}+\ldots, &  & \sigma\ll1,\\
\\
1-\frac{l\left(l+1\right)}{4\sigma}+\ldots, &  & \sigma\gg1.
\end{array}\right.\label{eq:AnisotropyFunction}
\end{equation}

For the calculation of the field and anisotropy asymptotes, we seek
an expansion of the magnetization in the low-field regime as 

\begin{eqnarray}
m & \simeq & \chi^{\left(1\right)}x+\chi^{\left(3\right)}x^{3}\label{eq:LowHDDIMag}
\end{eqnarray}
where \cite{jongar01prb} the coefficient of the linear contribution
reads
\begin{equation}
\chi^{\left(1\right)}=a_{0}^{\left(1\right)}+a_{1}^{\left(1\right)}\xi+a_{2}^{\left(1\right)}\xi^{2}\label{eq:MagChi1}
\end{equation}
with
\begin{eqnarray}
a_{0}^{\left(1\right)} & = & \frac{1+2S_{2}}{3},\quad a_{1}^{\left(1\right)}=a_{0}^{2}\mathcal{C}^{\left(0,0\right)},\nonumber \\
a_{2}^{\left(1\right)} & = & -\frac{2a_{0}^{2}}{3}\left[\left(1-S_{2}\right)\left(\bar{\mathcal{R}}-\mathcal{S}\right)+3S_{2}\left(\mathcal{T}-\mathcal{U}\right)\right]\nonumber \\
 &  & +\frac{b_{0}}{2}\left[\left(1-S_{2}\right)\mathcal{V}+3S_{2}\left(\mathcal{T}-\frac{1}{3}\bar{\mathcal{R}}\right)\right],\nonumber \\
\nonumber \\
b_{0} & = & \frac{4}{315}\left(7+10S_{2}-35S_{2}^{2}+18S_{4}\right).\label{eq:ab coefs}
\end{eqnarray}
Here, $\mathcal{C}^{\left(0,0\right)},\bar{\mathcal{R}},\mathcal{S},\mathcal{T},\mathcal{U},\mathcal{V}$
are certain lattice sums discussed in Ref.   \onlinecite{jongar01prb}.

The coefficient of the cubic contribution is given here only up to
first order in $\xi$

\begin{equation}
\chi^{\left(3\right)}=a_{0}^{\left(3\right)}\left[1+\xi4\mathcal{C}^{\left(0,0\right)}a_{0}^{\left(1\right)}\right]\label{eq:MagChi2}
\end{equation}
where
\[
a_{0}^{\left(3\right)}=-\frac{1}{315}\left(7+40S_{2}+70S_{2}^{2}-12S_{4}\right).
\]

Inserting the low- and high-field anisotropy expressions for $S_{l}\left(\sigma\right)$
into the coefficients of Eq. (\ref{eq:ab coefs}) and then substituting
the corresponding low- and high-field expansions of the Langevin function,
leads to the asymptotic equations of the main text. 

\bibliography{hkbib,/usr/share/texmf-texlive/bibtex/bib/hk/hkbib}

\end{document}